\documentclass[12pt]{article}
\usepackage[utf8]{inputenc}
\usepackage{amsthm,amssymb,amsmath,amsfonts,amstext,mathtools}
\usepackage{tikz}
\usetikzlibrary{shapes.geometric, arrows}
\usepackage{soul}
\usepackage{algpseudocode}
\usepackage{xr}

\tikzstyle{stop} = [rectangle, rounded corners,  minimum width=1.2cm, text width=1.2cm, minimum height=0.7cm, text centered, draw=cyan, fill=cyan!20]

\tikzstyle{stop2} = [rectangle, rounded corners,  minimum width=1.55cm, minimum height=0.7cm, text centered, draw=cyan, fill=cyan!20]

\tikzstyle{process} = [rectangle, rounded corners,  minimum width=1cm, minimum height=0.7cm, text centered, draw=cyan, fill=cyan!20]

\tikzstyle{decision} = [ellipse, minimum width=2cm, minimum height=0.7cm, text centered, fill=magenta!50]
\tikzstyle{arrow} = [thick, ->, >=stealth, draw=cyan, line width=1.2pt]

\usepackage[english]{babel}
\usepackage{array,latexsym}
\usepackage{xcolor}
\usepackage{graphicx,lscape}
\usepackage{subcaption}  

\usepackage{appendix}

\usepackage{floatrow}
\newfloatcommand{capbtabbox}{table}[][\FBwidth]
\newcommand{\blind}{0}
\usepackage{algpseudocode}
\usepackage{algorithm}

\usepackage{commath,mathrsfs,float,bbm,bm,dsfont,soul,tikz}
\usepackage{setspace,newlfont,authblk,lscape,longtable,multirow,comment,xfrac,indentfirst}
\usepackage[T1]{fontenc}
\usepackage[round]{natbib}
\usepackage[colorlinks=true, linkcolor=blue, citecolor=blue, filecolor=blue, urlcolor=blue, breaklinks=true] {hyperref}

\newcolumntype{G}{> {\columncolor[gray]{0.8}}c}

\setstcolor{red}
\newtheorem{proposition}{Proposition}

\newtheorem{lemma}{Lemma}

\newtheorem{corollary}{Corollary}
\newtheorem{assumption}{Assumption}
\newtheorem{remark}{Remark}
\newtheorem{example}{Example}
\date{}

\begin{document}

	\def\spacingset#1{\renewcommand{\baselinestretch}%
		{#1}\small\normalsize} \spacingset{1}
	\if0\blind
	{



\title{\bf The impact of job stability on monetary poverty in Italy: causal small area estimation}

\author{Katarzyna Reluga\footnote{University of Bristol, School of Mathematics, Bristol, United Kingdom, katarzyna.reluga@bristol.ac.uk}, Dehan Kong\footnote{Department of Statistical Sciences, University of Toronto, Canada, dehan.kong@utoronto.ca}, Setareh Ranjbar\footnote{Department of Psychiatry, Lausanne University Hospital, University of Lausanne, Switzerland, setareh.ranjbar@chuv.ch}, Nicola Salvati\footnote{Department of Economics and Management, University of Pisa, Italy, nicola.salvati@unipi.it}, Mark van der Laan\footnote{School of Public Health, University of California, Berkeley, USA, laan@berkeley.edu\\ {{ The authors gratefully acknowledge support from the Swiss National Science Foundation for the project P2GEP2-195898 and the Center for Targeted Machine Learning, University of Califronia at Berkeley. \\The computations were performed at the University of Geneva using Baobab and Yggdrasil HPC Service and using     the computational facilities of the Advanced Computing Research Centre, University of Bristol.}}}
		}
  
		\maketitle
	} \fi	
	
	\if1\blind
	{\bigskip
		\bigskip
		\bigskip
		\begin{center}
			{\LARGE\bf The impact of job stability on monetary poverty in Italy: causal small area estimation\par}
		\end{center}
	}  \fi
\vspace{-1cm}
	\begin{abstract}
		\noindent
Job stability -- encompassing secure contracts, adequate wages, social benefits, and career opportunities --  is a critical determinant in reducing monetary poverty, as it provides households with reliable income and enhances economic well-being. This study leverages EU-SILC survey and census data to estimate the causal effect of job stability on monetary poverty across Italian provinces, quantifying its influence and analyzing regional disparities. We introduce a novel causal small area estimation (CSAE) framework that integrates global and local estimation strategies for heterogeneous treatment effect estimation, effectively addressing data sparsity at the provincial level. Furthermore, we develop a general bootstrap scheme to construct reliable confidence intervals,  applicable regardless of the method used for estimating nuisance parameters. Extensive simulation studies demonstrate that our proposed estimators outperform classical causal inference methods in terms of stability while maintaining computational scalability for large datasets. Applying this methodology to real-world data, we uncover significant relationships between job stability and poverty across six Italian regions, offering critical insights into regional disparities and their implications for evidence-based policy design.

\end{abstract}
	
	\noindent%
	
	\vfill
	
	\newpage
	\spacingset{1.45} 
 
\section{Introduction}
\subsection{Relationship of job stability and  monetary poverty in Italy}\label{sec:impact_job}
To what extent does job stability influence monetary poverty? This question is relevant to both labour economists and policymakers. Job stability, encompassing secure contracts, adequate wages, social benefits, and career development, is essential for reducing poverty by ensuring consistent income and enhancing economic well-being \citep{temporary_jobs}. 

The influence of stable employment varies by region, shaped by economic and social conditions such as local labor markets. In Italy, pronounced north-south socioeconomic disparities persist \citep{italy_divide}, stemming from the South’s ongoing difficulty in attracting businesses and creating jobs. In this context, accurate estimates of the relationship between job stability and monetary poverty across Italian provinces can guide policymakers in designing targeted interventions to reduce territorial inequalities.

Scholars have extensively studied the link between job stability and poverty in Italy. \citet{Mussida2021} emphasize the importance of stable employment and social protection in mitigating monetary poverty, particularly in high-unemployment regions. Focusing on Italian macro-regions, they uncover evidence of strong state dependence and increasing scarring effects in southern Italy. \citet{Gallo2018} show how job instability, including long-term unemployment, harms future employment prospects and earnings, increasing poverty risk. Similarly, \citet{filandri2020being} find that temporary contracts heighten subjective poverty, regardless of household income.

Previous studies have primarily explored the relationship between job stability and poverty at the macro-regional level, without employing causal inference methods or addressing smaller-scale heterogeneity. Yet, assessing this effect at provincial and municipal levels is crucial, as intra-regional disparities can exceed inter-regional ones. Local policymakers, tasked with reducing poverty and improving socioeconomic conditions, can use insights into this relationship to design targeted strategies such as job creation, skills development, and fostering a favourable business environment. These efforts not only tackle regional inequalities but also enhance the well-being of local communities.

A plausible approach to assessing the relationship between job stability and monetary poverty is to examine the effect of a variable indicative of job stability, such as contract type, on a monetary poverty indicator like household equivalised income. For instance, one could compare households where the head has a permanent contract to those with a short-term contract, examining how these differences vary across Italian provinces. Treating contract type as an intervention variable, 
these differences can be considered naive estimates of the average treatment effects  \citep[ATE,][]{Imbens2015}. 

\begin{figure}[ht]
\centering 
\includegraphics[width=0.32\textwidth]{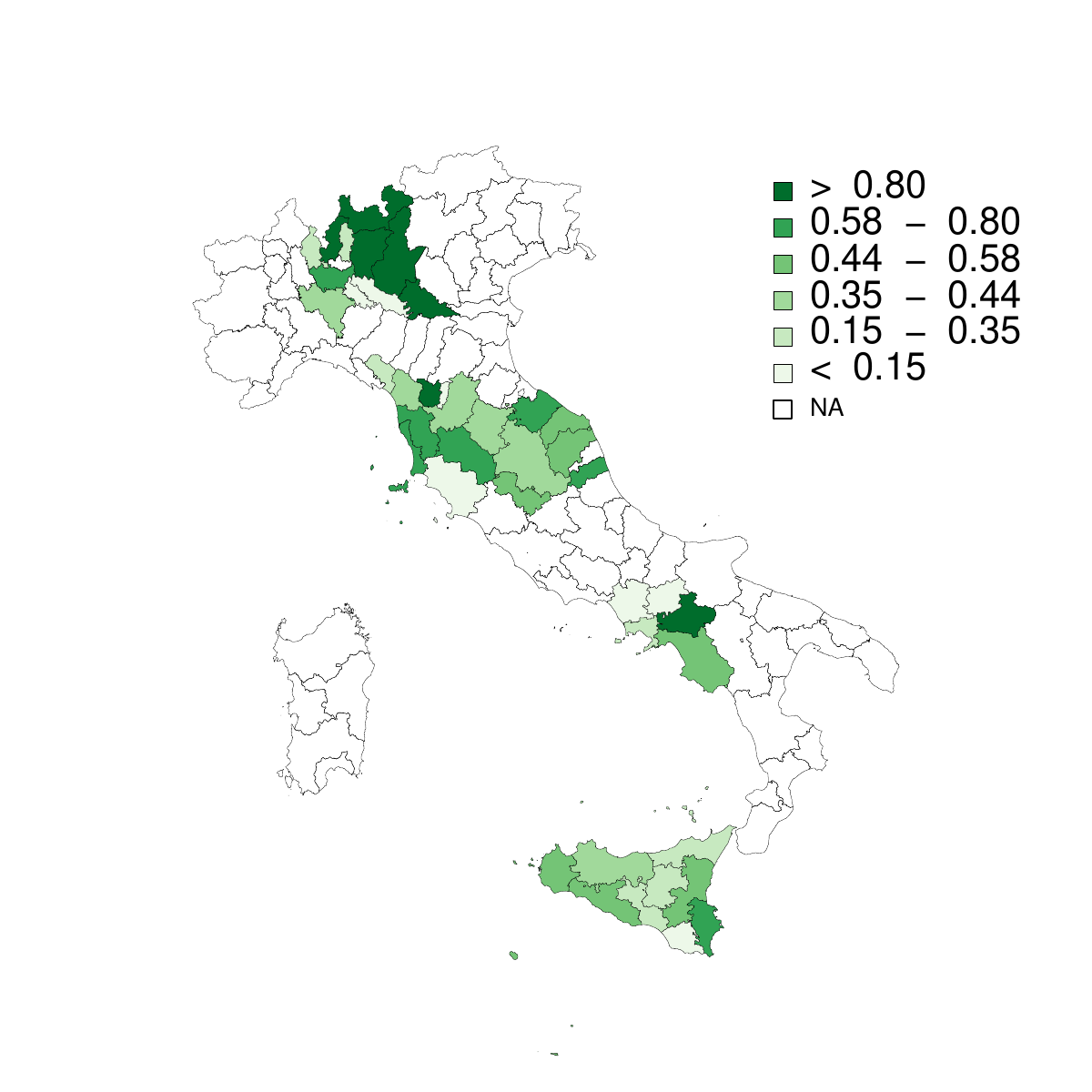} 
\includegraphics[width=0.32\textwidth]{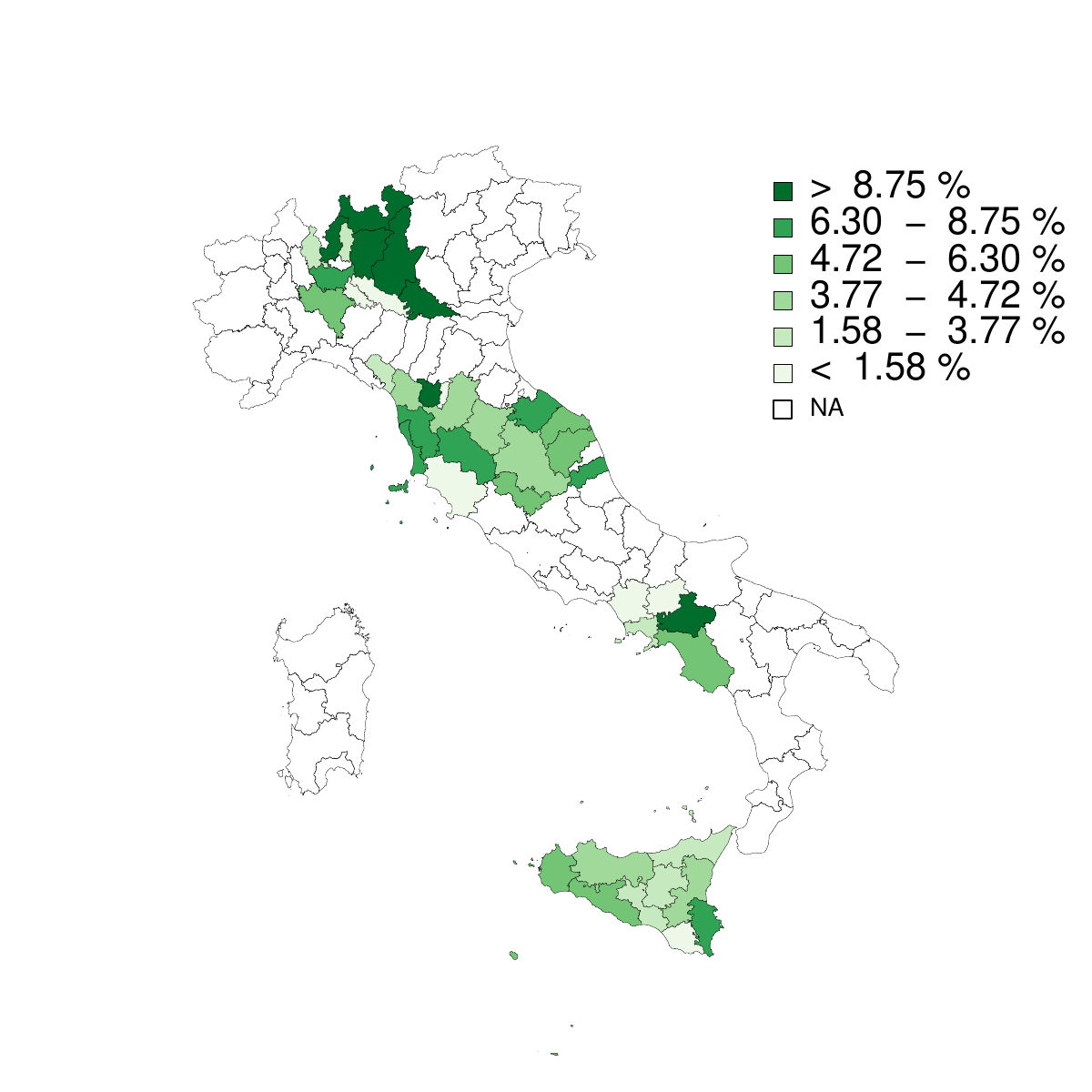}   \includegraphics[width=0.32\textwidth]{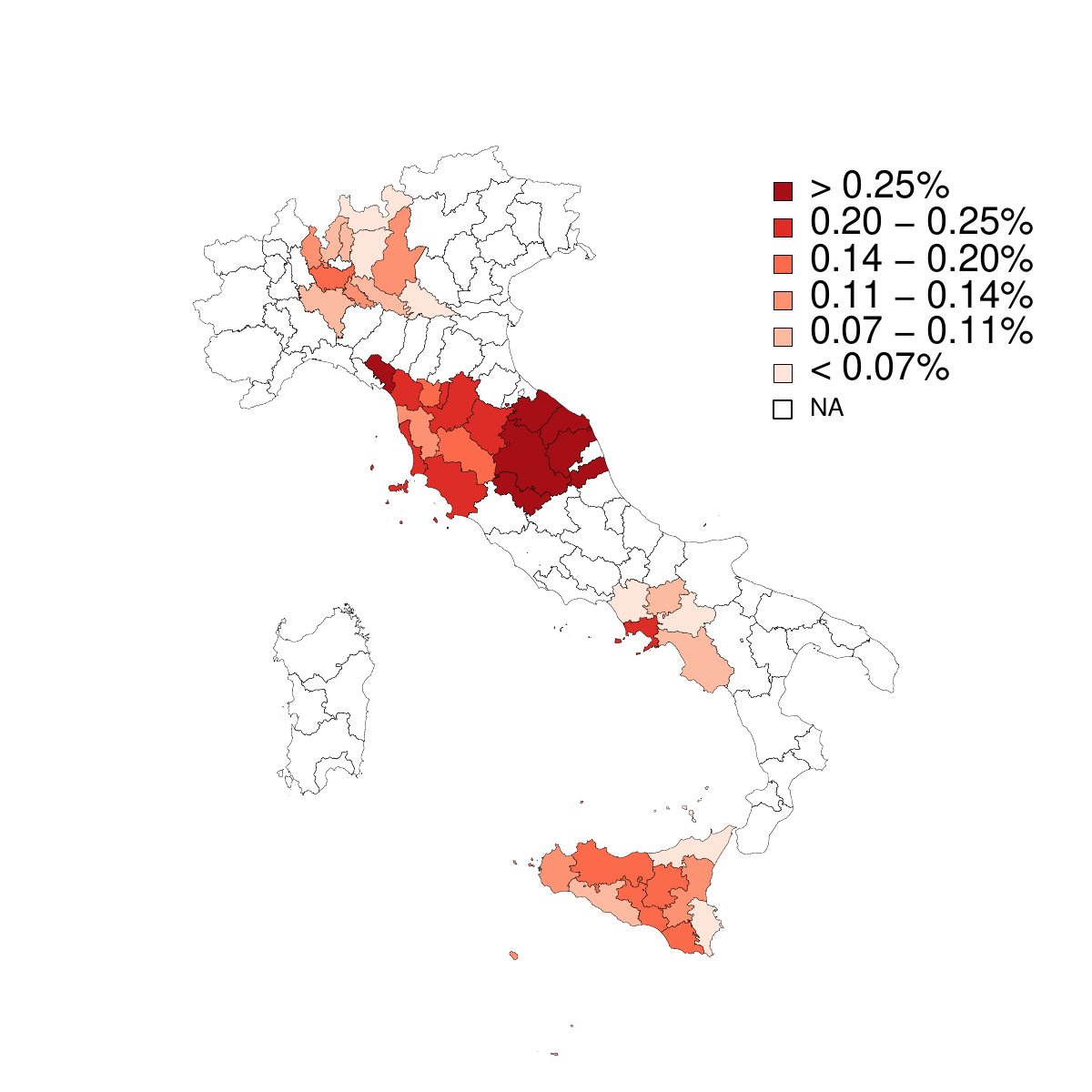}
\caption{The maps of naive direct estimates of ATE on an original scale (left) and in \% (middle); 
the sampling fraction for the provinces of six regions of Italy (right).}
\label{fig:map_comp}
\end{figure}

\begin{figure}[ht]
		\centering \includegraphics[width=0.8\textwidth]{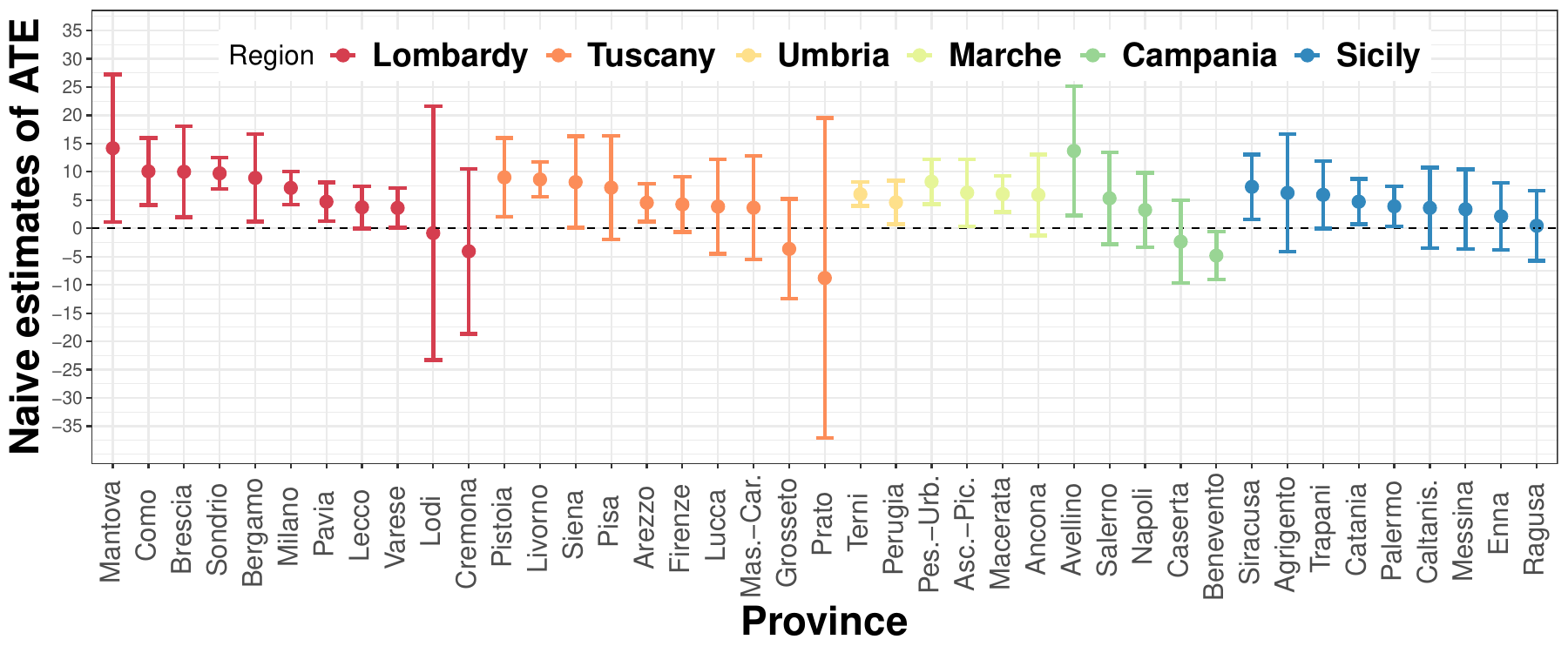} 
		\caption{The confidence intervals of the naive ATE estimates in \% (represented with dots).}
		\label{fig:sate}
	\end{figure}

The left panel of Figure~\ref{fig:map_comp} shows differences in direct Hajek estimates \citep{hajek1971comment} of log equivalised income between treated and control households, whereas the middle panel presents the naive ATE as a percentage ratio of the average income of treated to control households (see Example~\ref{example:did} and Secion~\ref{sec:data_analysis_impact} for a formal definitions). Even though the estimates in both panels are biased due to confounding in observational data, and they are intended for illustrative purposes only, using data from the 2012 EU-SILC survey (Section \ref{sec:description of data}), we observe substantial heterogeneity in the relationship between job stability and poverty across provinces, with permanent contracts associated with changes in equivalised income ranging from -8.78\% to 14.19\%. 
At the same time, Figure \ref{fig:sate} presents the naive estimates of ATE together with their confidence intervals. The width of intervals for many provinces does not allow for establishing a priority order for policymakers or drawing firm conclusions about the relationship between job stability and monetary poverty. This is explained by the right panel of Figure \ref{fig:map_comp}, which depicts the sampling fraction -- ranging from 0.02\% to 0.53\% -- indicating precariously small sample sizes. Such sparsity makes conventional direct estimators unsuitable for evaluating the effect of job stability on monetary poverty at the provincial level.
When the sample size for an area of interest is small or zero, researchers face the small area estimation (SAE) dilemma.  To our knowledge, apart from a recent proposal of \cite{ranjbar2021causal} which is a special example of our general framework, the existing causal inference methods are unsuitable for our case study in which only a tiny fraction of outcomes in each province is actually observed. We propose alternative statistical methods to derive meaningful conclusions.

\subsection{Literature review}\label{sec:literature_review}

This section outlines key elements of the literature relevant to our work, focusing on five strands: (a) measuring local poverty rates in survey studies, (b) assessing treatment-effect heterogeneity in observational studies, (c) semi-supervised inference for treatment effects, (d) causal inference in SAE, and (e) implementing machine learning methods in SAE.

\textbf{Measuring poverty at the small area level}. SAE combines survey samples with auxiliary data to provide accurate statistics for subpopulations with small sample sizes \citep[][]{rao2015small,morales2021course}. The developments in SAE have been driven by, among others, an interest in providing reliable estimates of local poverty rates. Various authors 
proposed model-based parametric estimation within frequentist or Bayesian framework, including \cite{aosMolina,marhuenda} and a monograph of \cite{pratesi2016analysis}. However, none of these authors considered poverty estimation within a causal framework, which is the focus of our study.

\textbf{Assessing treatment-effect heterogeneity in observational studies}. There exists an extensive literature on heterogeneous treatment-effect estimation in observational studies; recent discussions on this topic can be found in, among others, 
\cite{anoke2019approaches}.
Most popular estimators involve fitting a model for the outcome surface, such as outcome regression (OR) estimators 
\citep[][]{Imbens2015}; modelling the treatment assignment mechanism, i.e., inverse propensity weighting (IPW) estimators \citep{horvitz,hajek1971comment};  or modeling both jointly through augmented IPW (AIPW) estimators \citep{robins1994estimation}, 
targeted maximum likelihood estimators \citep[TMLE][]{van2011targeted}, or debiased machine learning (DML) estimator \citep{chernozhukov2018}. In the context of multilevel and clustered data, IPW-type estimators have proven successful in both frequentist and Bayesian setups 
\citep{zubizarreta2017optimal,lee2021partially}.
However, the above methods cannot be directly applied to our setup due to the extreme scarcity of data at the level of subpopulations and the need to pool information from other subpopulations in order to improve the efficiency of new estimators.

\textbf{Semi-supervised inference for treatment effect}. Our problem setup resembles semi-supervised inference for treatment effects, where only a small fraction of the data is labeled, and for the remaining units, only covariates and surrogates are available \citep{cheng2021robust,kallus2022role}. However, unlike these studies, we lack high-quality surrogates for the unlabeled subpopulations and face a more extreme type of estimation in which the fraction of labelled data is minimal. Directly applying their methods would result in estimates with excessive mean squared error (MSE).

\textbf{Causal inference for SAE}.
Causal inference in SAE has received little attention until the works of \cite{ranjbar2021causal} and \cite{gao2022treatment}. \cite{ranjbar2021causal} proposed an IPW-type estimator for heterogeneous treatment effects, using parametric techniques like mixed models and M-quantile regression to estimate nuisance parameters and establish asymptotic properties. Their method is a specific instance of our broader framework. In contrast, \cite{gao2022treatment} applied Bayesian multilevel regression and post-stratification to estimate ATE and conditional ATE in randomised experiments, which differs from our focus on observational data.

\textbf{ML in SAE}. The integration of ML in SAE has gained traction, particularly with tree-based methods used for, among others, poverty estimation in Nepal \citep{bilton2017classification}, 
and household income in Mexico \citep{krennmair2022flexible}. However, these methods have not yet addressed causal estimation. Our study employs a broad library of ML and parametric methods to estimate nuisance parameters, which play a critical role in determining the performance of the estimators.

\subsection{Our contribution}\label{sec:contribution}	
To address the empirical challenges outlined in Section~\ref{sec:impact_job}, we propose a causal small area estimation (CSAE) framework to study the impact of job stability on monetary poverty across Italian provinces. Our new estimation methods aim to reliably quantify this relationship while investigating heterogeneity both between and within Italian regions. Our CSAE framework is specifically designed to: (a) address the constraints of the case study, particularly the negligible fraction of observed outcomes at the subpopulation level; (b) leverage SAE principles to pool information across provinces, thus improving the precision of the estimates; and (c) integrate efficient estimators by jointly modelling outcome regression and treatment assignment. To the best of our knowledge, no existing method simultaneously meets these three criteria.

Our CSAE framework incorporates global and local estimators, extending classical OR-type, IPW-type, and AIPW-type methods to address outcome missingness due to sampling. We show that classical doubly robust estimators are unattainable in this context and outline the asymptotic normality of our proposed estimators. Additionally, we develop a consistent bootstrap scheme for constructing reliable confidence intervals, regardless of the nuisance parameter estimation method. This work bridges observational causal inference and SAE  by examining the underlying assumptions governing both fields. We conducted an extensive sensitivity analysis to evaluate our estimators against classical methods using simulations based on a synthetic population emulating the case study. The simulations, which were unbiased toward any specific estimation method, showed that the choice of strategy for predicting out-of-sample outcomes is critical to estimator performance -- aligning with existing SAE literature.  We found that the AIPW-type estimator outperformed competing methods due to its stability and computational scalability for large datasets. In our case study, the AIPW-type estimator revealed that job stability affects monetary poverty more within regions than across regions in Italy, a conclusion unattainable with regional-level data alone. The positive effect of job stability on reducing poverty is stronger in northern Italy, where labour markets are more regulated and industrial development is higher. The analysis and numerical experiments were conducted using the publicly available R package \texttt{causalSAE} \citep{causalSAE}. Substantively, our work contributes to the poverty mapping literature by offering new insights into the variability of how job stability affects monetary poverty across provinces with differing socio-economic conditions.

The manuscript is organized as follows. Section~\ref{sec:description of data} describes the case study data, while Section~\ref{sec:csae} outlines the CSAE framework and identification arguments. Section~\ref{sec:methodology} introduces a new class of estimators, and Section~\ref{sec:boot_theory} details the bootstrap scheme for asymptotically valid confidence intervals. Section~\ref{sec:simulations} presents a sensitivity analysis via simulations, and Section~\ref{sec:data_analysis_impact} applies these results to assess the impact of job stability on relative poverty in Italian provinces. Section~\ref{sec:discussion} discusses future research directions. Appendix includes proofs of 
asymptotic normality of new estimators and consistency of bootstrap confidence intervals as we as additional results from simulations and the data analysis.


\section{Socio-economic survey and census data in Italy}\label{sec:description of data}
The case study used two data sources: administrative data from the 2011 Italian Census and survey data from the 2012 EU-SILC, making 2011 the reference year. The Italian National Institute of Statistics (ISTAT) conducts EU-SILC annually to produce living conditions indicators at national and regional (NUTS 2) levels, leveraging both datasets for statistical analyses. EU-SILC employs a stratified two-stage sampling design, with municipalities as primary sampling units and households as secondary units. For details on the design, see \cite{ISTAT_EU_SILC}. In our analysis, we account for the two-stage sampling design by evaluating outcomes at the household level and using regions as covariates. 
Due to limited sample sizes at the municipal level, we focus on provinces and households, a widely accepted approximation of EU-SILC’s two-stage design within the SAE community.  

Our study examines the impact of job stability on monetary poverty across 41 Italian provinces, referred to as subpopulations or areas, spanning six regions: Lombardy (11), Tuscany (10), Umbria (2), Marche (4), Campania (5), and Sicily (8). These regions represent northern, central, southern, and insular Italy, allowing for a comprehensive analysis of the north-south divide. The effect of job stability on poverty alleviation is expected to be stronger in central and northern regions, where economies are more stable (see Figure~\ref{fig:map_comp}).

\begin{figure}[ht]
\centering 
\includegraphics[width=0.31\textwidth]{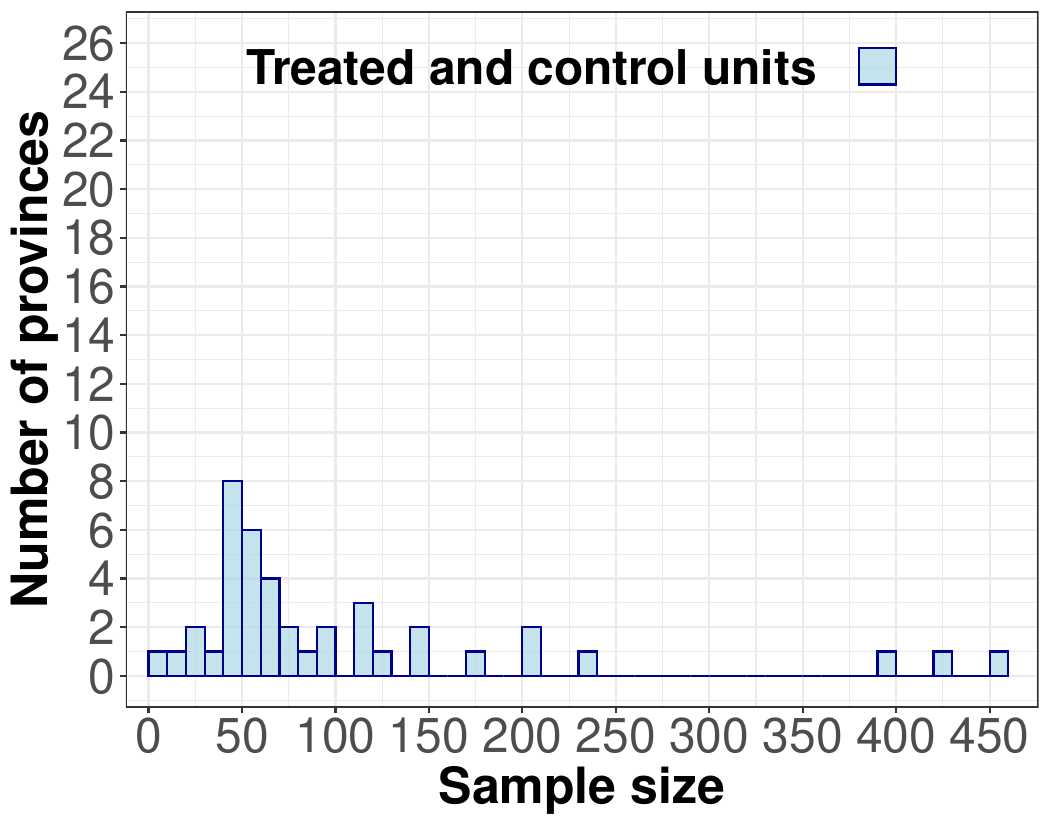} 
\includegraphics[width=0.31\textwidth]{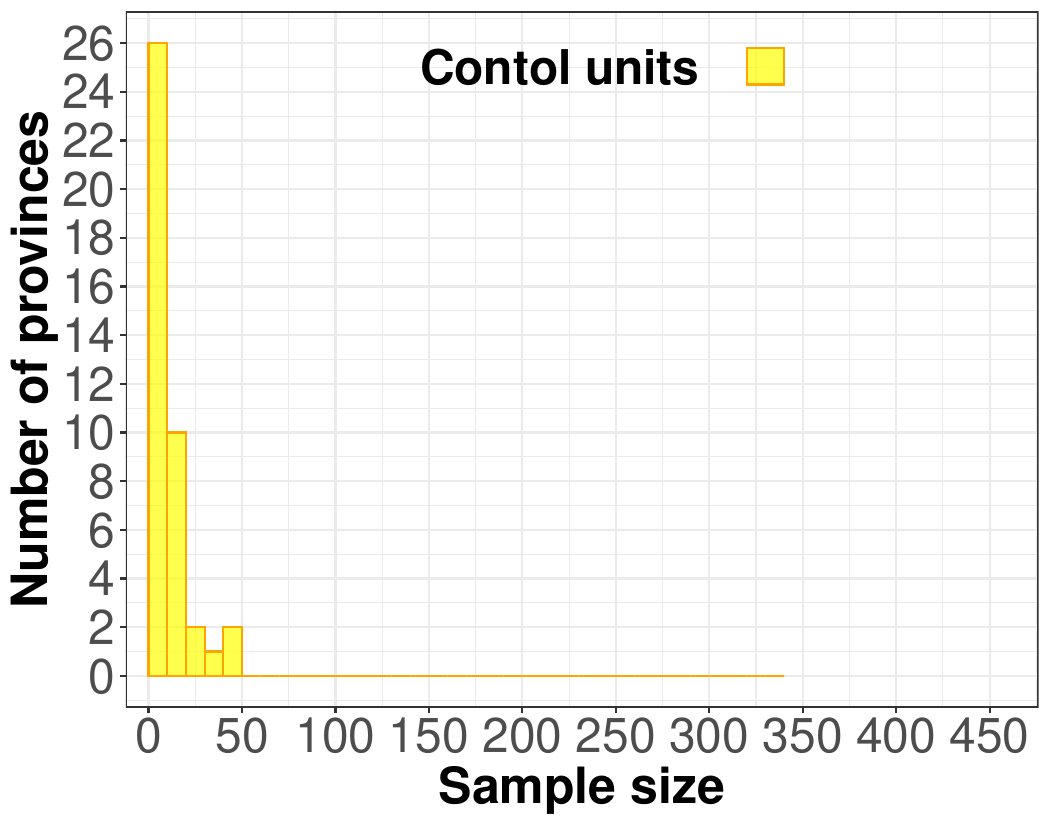}   \includegraphics[width=0.31\textwidth]{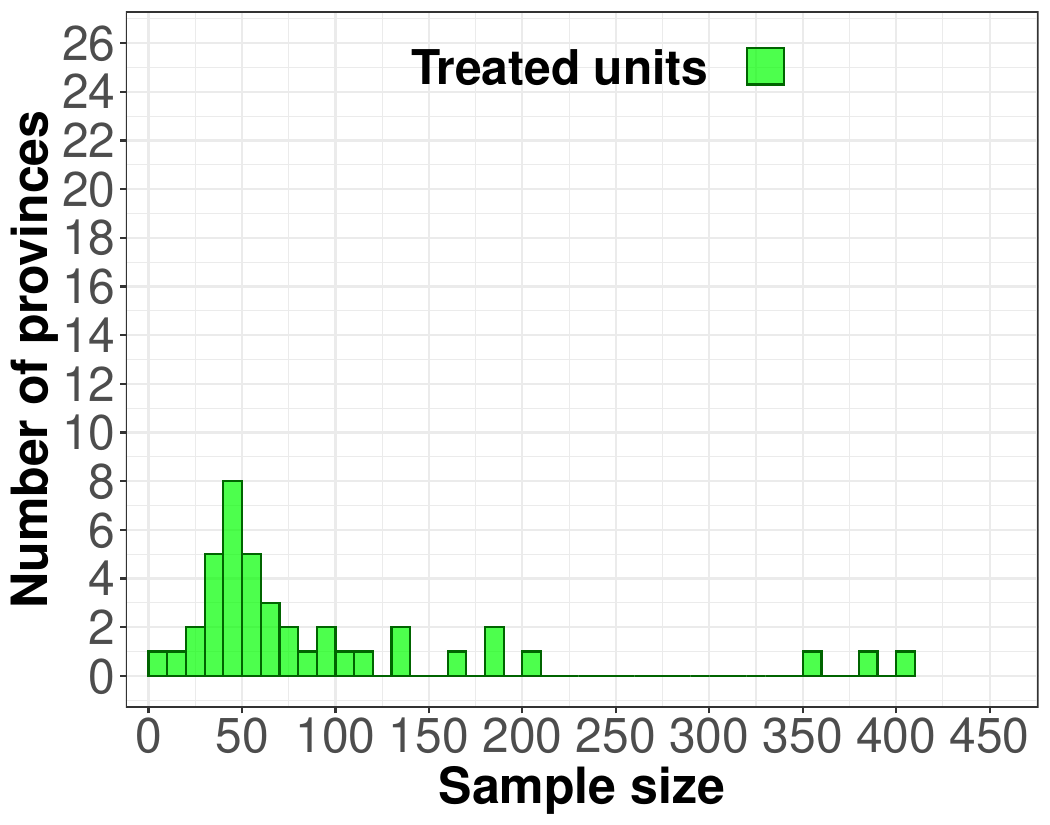}
	\caption{Distribution of the sample size by province and by treatment status.}
	\label{fig:SampleSize}
 \end{figure}

The total population size is \( N = 2,874,217 \), with a sample size of \( n = 4,371 \) and a sampling fraction of \( f = n/N \approx 0.0015 \). Among observed provinces, sample sizes range from 9 to 454 (mean: 106, median: 64). A substantial imbalance exists between treated and control groups. For household heads with permanent contracts, sample sizes range from 8 to 404 (mean: 95, median: 58), while for those with temporary contracts, they range from 1 to 50 (mean: 11, median: 7). Figure~\ref{fig:SampleSize} presents the histogram of provinces by sample size. The presence of many provinces with very small samples (see also the third panel of Figure~\ref{fig:map_comp}) hinders reliable direct estimation at the area level, reinforcing the need for SAE techniques.

We use the household head’s contract type as an indicator of job stability and measure monetary poverty via log equivalised disposable income. The latter is computed by dividing total household disposable income by a factor accounting for household size and composition, based on the modified OECD scale \citep{Hagenaars:1994}. Population-level covariates come from the 2011 census, while variable descriptions from EU-SILC 2012 and the 2011 census are provided in Section~\ref{sec:data_analysis_impact}. Some variables pertain to the household head, others to the household level.

\section{Basic setup for causal small area estimation}\label{sec:csae}
Consider a population of Italian regions $\mathcal{U}$ of size $N$, partitioned into $m$ mutually disjoint provinces (subpopulations/areas) $\mathcal{U}_{j}$ of size $N_{j}$, such that $\mathcal{U} = \bigcup_{j\in \mathcal{V}}^{}\mathcal{U}_j$, $\mathcal{V}=\{1, \dots, m\}$. Each subpopulation $\mathcal{U}_j$ is further partitioned into a sampled part, $\mathcal{U}^s_j = \{1, \dots, n_j\}$, and a remaining part, $\mathcal{U}^r_j = \{n_j+1, \dots, N_j\}$, with sizes $n_j$ and $N_j - n_j$, respectively, with $N_j \gg n_j$. We have the following relationships:
\begin{equation}\label{eq:sizes}
       n = \sum_{j=1}^{m} n_j, \quad N = \sum_{j=1}^{m} N_j, \quad \mathcal{U}^s = \bigcup_{j=1}^{m} \mathcal{U}^s_j, \quad \mathcal{U}^r = \bigcup_{j=1}^{m} \mathcal{U}^r_j, \quad \mathcal{U} = \mathcal{U}^r \cup \mathcal{U}^s.
\end{equation}

\begin{figure}[ht]
			\begin{floatrow}
				\capbtabbox{%
					\begin{tabular}{c c c c c c}
						$\bm{X}_{ij}$ & $A_{ij}$ & $S_{ij}$ & $Y_{ij}$  & {\color{gray}$Y_{ij}(1)$} & {\color{gray}$Y_{ij}(0)$} \\
						\hline
						$\bm{X}_{1j}$ &1 & 1 & $Y_{1j}$  & {\color{gray}$Y_{1j}(1)$} & {\color{gray}NA}  \\
						$\bm{X}_{2j}$ &0 & 1 & $Y_{2j}$  & {\color{gray}NA} & {\color{gray}$Y_{2j}(0)$}  \\
						$\vdots$ &$\vdots$& $\vdots$ & $\vdots$  & {\color{gray}$\vdots$} & {\color{gray}$\vdots$}  \\
						$\bm{X}_{n_jj}$ &0 & 1 & $Y_{n_jj}$  & {\color{gray}NA} & {\color{gray}$Y_{n_jj}(0)$}  \\
						$\bm{X}_{(n_j+1)j}$ &1 & 0 & NA  & {\color{gray}NA} & {\color{gray}NA}  \\
						$\vdots$ &$\vdots$& $\vdots$ & $\vdots$  & {\color{gray}$\vdots$} & {\color{gray}$\vdots$}  \\
						$\bm{X}_{(N_j-1)j}$ &1 & 0 & NA  & {\color{gray}NA} & {\color{gray}NA}  \\
						$\bm{X}_{N_jj}$ &0 & 0 & NA  & {\color{gray}NA} & {\color{gray}NA}  \\\hline
					\end{tabular}
				}{%
					\caption{Schematic representation of data at the subpopulation-level $j\in \mathcal{V}_j$. \label{tab:data}}
				}%
				\ffigbox{%
					\includegraphics[width=0.3\textwidth]{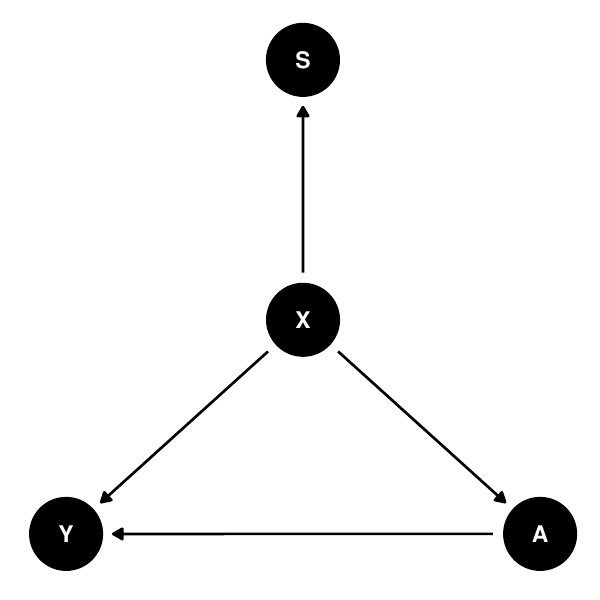}%
				}{%
					\caption{Direct acyclic graph of the data.  \label{fig:dag}}%
				}
			\end{floatrow}
		\end{figure}

Let $A_{ij}\in \{0, 1\}$ denote a binary treatment indicator, where $A = 1$ for treated units and $A = 0$ for controls. Let $S_{ij}\in \{0, 1\}$ be a sample membership indicator, where $S_{ij}=1$ if $Y_{ij}\in \mathcal{U}^s$ and $S_{ij}=0$ otherwise. 
Define \(\bm{X}_{ij} = (\bm{W}^T_{ij}, G_{ij})^T = (W_{ij1}, W_{ij2}, \dots, _{ijq}, G_{ij})^T \in \mathcal{X} \subset \mathbb{R}^{q+1}\) as a vector of covariates or auxiliary variables, where \(G_{ij} \in \mathcal{V}\) denotes the subpopulation indicator. Without loss of generality, we assume that the first \( p \) covariates of \( \bm{W}_{ij} \) are individual-level, meaning \( W_{ijk} \neq W_{i'j'k} \) unless \( i = i' \), \( j = j' \), \( k = k' \) for \( k = 1, \dots, p \). The remaining \( q - p \) covariates, commonly referred to as contextual variables \citep{Lyu:2022}, are subpopulation-level, meaning \( W_{ijl} = W_{i'j'l} \) for all \( i, i' \in \mathcal{U}_j \) when \( j = j' \) and \( l = p+1, \dots, q \).  Thus, the data contain information at both individual and subpopulation levels. Let \( Y_{ij} \in \mathbb{R} \) denote the observed outcome, defined as:
\begin{equation} \label{eq:causal_cons}
    Y_{ij} = Y_{ij}(1) A_{ij} + Y_{ij}(0) (1 - A_{ij}) \quad \text{when} \quad S_{ij} \in \{0,1\}.
\end{equation}
Here, $ Y_{ij}(a) $ represents the potential (counterfactual) outcome of a unit under treatment status \( A_{ij} = a \) for \( a \in \{0,1\} \). The definition of \( Y_{ij} \) implies the classical consistency assumption, which is necessary for identifying the causal parameter in \eqref{eq:iden_all}.  
Finally, let  \( \bm{Z}^T_{ij} = (Y_{ij}, \bm{X}_{ij}^T, A_{ij})^T \) for \( i \in \mathcal{U}^s_j \), and \( \bm{X}^T_{A,ij} = (\bm{X}_{ij}^T, A_{ij})^T \) for \( i \in \mathcal{U}^r_j \), with \( \bm{Z}_{ij} \sim P_j \) and \( \bm{X}_{Aij} \sim P_j^{\bm{X}_A} \), representing the true data distributions of \( \bm{Z}_{ij} \) and \( \bm{X}_{A,ij} \), respectively, within the subpopulation \( j \in \mathcal{V} \). Figure \ref{fig:dag} and Table \ref{tab:data} show a schematic representation of the 
data. The causal estimand of interest is the subpopulation-level average treatment effect:
\begin{equation}\label{eq:param_int}
\tau_j=\tau_j(1) - \tau_j(0)= E\{Y_{ij}(1)|G_{ij}= j\} - E\{Y_{ij}(0)|G_{ij}= j\} =  E_j\{Y_{ij}(1)\} - E_j\{Y_{ij}(0)\}.
\end{equation}
To guarantee that $\tau_j$ in \eqref{eq:param_int} is identifiable and estimable from the available data, we adopt following assumptions. 

\begin{assumption}\label{assump:key_sae1} 
$\bm{Z}_{ij}$ and $\bm{X}_{A, ij}$ share the same underlying distribution in the following sense
$P^{\bm{X}_A}_j(\bm{X}_{A, ij})  = \int P_j(\bm{Z}_{ij})dY_{ij}$. Thus, $\{\bm{Z}_{ij}^T: i \in \mathcal{U}^s \}$ and $\{\bm{X}^T_{A,ij}: i \in \mathcal{U}^r\}$ are realizations from $P_j$ and $P_j^{\bm{X}_A}$, respectively, 
for all $j \in \mathcal{V}$.
\end{assumption}

\begin{assumption}[Unconfoundeness in SAE]\label{assump:ignorability} Let $ a = 0,1 $. Then $ A_{ij} \perp Y_{ij}(a)|\bm{X}_{ij}, S_{ij} \text{ and } S_{ij} \perp Y_{ij}(a), A_{ij} |\bm{X}_{ij}$.
\end{assumption}

\begin{assumption}[Overlap in SAE]\label{assump:overlap} Let $ e_1(\bm{x}_{ij})\coloneqq P(A_{ij}=1|\bm{X}_{ij} = \bm{x}_{ij})$ and $e_0(x_{ij}) = 1-e_1(\bm{x}_{ij})$. For any $x_{ij}\in\mathcal{X}$ and some constants $c_1, c_2\in (0,1)$,  we have $ e_1(\bm{x}_{ij})\in (c_1, 1-c_2) $.
\end{assumption}

Assumption \ref{assump:key_sae1} underpins the SAE paradigm called a \textit{predictive approach} \citep{valliant2000finite, morales2021course}, which posits a superpopulation (or working) model applicable across all subpopulations $\mathcal{U}_j$ for both sampled and remaining units. Here, we frame this in terms of a data generation process rather than a model. Additionally, the first component of Assumption \ref{assump:ignorability} extends the no unmeasured confounders assumption \citep{rubin1978bayesian}, while the second ensures the absence of selection bias (or noninformative sampling), a key assumption in SAE. Under Assumption \ref{assump:ignorability}, it follows that:
\begin{align}
\mu(\bm{X}_{A,ij}) &= E(Y_{ij}|\bm{X}_{A,ij}, S_{ij} = 1) = E(Y_{ij}|\bm{X}_{A,ij}, S_{ij} = 0),\label{eq:mu}\\
\mu_a(\bm{X}_{ij}) &= E(Y_{ij}|A_{ij} = a, \bm{X}_{ij}, S_{ij} = 1) = E(Y_{ij}|A_{ij} = a, \bm{X}_{ij}, S_{ij} = 0),\label{eq:mu_a}\\
e_a(\bm{x}_{ij}) &= P(A_{ij} = a|\bm{X}_{ij} = \bm{x}_{ij}, S_{ij} = 1)= P(A_{ij} = a|\bm{X}_{ij} = \bm{x}_{ij}, S_{ij} = 0), a = 0, 1.\label{eq:ps_s01}
\end{align}
Assumptions~\eqref{assump:key_sae1}--\eqref{assump:overlap} ensure the identification of the causal estimand $\tau_j(a)$ in \eqref{eq:param_int} defined in terms of \textit{unobservable} potential outcomes. Identification is achieved 
through the statistical target parameter
$\tau_j^a$ expressed in terms of a random variable $\bm{Z}$ whose realisations are \textit{observable data}, that is 
\begin{align}\label{eq:iden_all}
\tau_j(a) &= E_j\{Y_{ij}(a)\} = 
E_j\{\mu_{a}(\bm{X}_{ij})\} = E_j\left\{\frac{I(A_{ij} = a)Y_{ij}}{e_a(\bm{X}_{ij})}\right\}\nonumber \\ &= E_j\left[
\frac{I(A_{ij} = a)}{e_a(\bm{X}_{ij})}\left\{Y_{ij}-\mu_a(\bm{X}_{ij})\right\} + \mu_a(\bm{X}_{ij})\right] \eqqcolon \tau_j^a.
\end{align}
The derivation of equivalence in~\eqref{eq:iden_all} relies on classical arguments and is thus deferred to the Appendix~\ref{appendix:iden}. In the following, we focus on constructing high-quality estimators for $\tau_j^a$ within the small area model-based framework.

\begin{remark}
Assumptions~\eqref{assump:key_sae1}--\eqref{assump:overlap}, along with causal consistency in \eqref{eq:causal_cons}, enable estimation methods targeting $\tau_j^{a}$ rather than the causal estimand $\tau_j(a)$. Thus, we do not assess the validity of the identifying assumptions, though their importance remains paramount.
\end{remark}

\begin{remark}
In survey statistics, an alternative parameter of interest is the finite-sample average treatment effect, $\tau_j^F=N_j^{-1}\sum_{i\in \mathcal{U}_j}\left\{Y_{ij}(1) - Y_{ij}(0)\right\}$. In randomized controlled trials, \cite{stuart2011use} proposed a Horvitz-Thompson-type estimator combining survey and propensity score weighting. However, extending this approach to our small-area setting is impractical because (a) our study is observational, and (b) the sample sizes $n_j$ are too small to ensure precise estimates at the subpopulation level $\mathcal{U}_j$ (cf. Figure \ref{fig:map_comp}).
\end{remark}

\vspace{-1cm}
\section{Estimation strategies}\label{sec:methodology}

\subsection{General linear weighting estimator}\label{sec:notation_est}

Before introducing a definition of a general causal small-area estimator, some additional notation is established. Let $\hat{\bm{\eta}}= \{\hat{\mu}_{n}(\cdot), \hat{\mu}_{a,n}(\cdot), \hat{e}_{1, N}(\cdot)\}$ be some estimators of $\eta= \{\mu(\cdot), \mu_{a}(\cdot), e_{1}(\cdot)\}$, whose elements are defined in \eqref{eq:mu} -- \eqref{eq:ps_s01}, and $\hat{e}_{0, N}(\cdot) = 1-\hat{e}_{1, N}(\cdot) $. Now, define  $\hat{Y}_{ij} = \hat{\mu}_n(\bm{X}_{A, ij})$  and \( \hat{\bm{Z}}^T_{ij} = (\hat{Y}_{ij}, \bm{X}_{ij}^T, A_{ij})^T \) for  $i\in \mathcal{U}^r_j$, and  
\(\hat{\bm{Z}}_{ij} = \bm{Z}_{ij}\) for $i\in \mathcal{U}^s_j$, $j\in\mathcal{V}$. For a random function \(\hat{f}(\cdot)\), define $\mathbb{P}_{N_j}\{\hat{f}(\hat{\bm{Z}}_{ij})\}  =\frac{1}{N_j} \sum_{j=1}^{N_j} \hat{f}(\hat{\bm{Z}}_{ij}), \;
\mathbb{P}_{\hat{N}^a_j}\{\hat{f}(\hat{\bm{Z}}_{ij})\} = \frac{1}{\hat{N}^a_j}\sum_{j=1}^{N_j} \hat{f}(\hat{\bm{Z}}_{ij})$ where $\hat{N}^a_j=\sum_{i=1}^{N_{j}}\frac{I(A_{ij} =a)}{\hat{e}_{a, N}(\bm{X}_{ij})}$. Let \(\varphi^a(\cdot, \hat{\bm{\eta}})\) be a function of the nuisance parameter $\hat{\bm{\eta}}$. A general linear weighting estimator of $\tau_j$ in \eqref{eq:param_int} is: 
\begin{equation} \label{eq:tau_hat}
\hat{\tau}_j = \hat{\tau}^1_j - \hat{\tau}^0_j =    
\mathbb{P}_{k}  \{\varphi^{1}(\hat{\bm{Z}}_{ij}; \hat{\eta})\} - \mathbb{P}_{k}  \{\varphi^{0}(\bm{\hat{Z}}_{ij};\hat{\bm{\eta}})\}, \; k \in \{N_j,\hat{N}^a_j\}.
\end{equation}
Estimators of the form given in \eqref{eq:tau_hat} are referred to as causal small area estimation (CSAE) techniques. This form aligns with the predictive approach in SAE (see Assumption~\ref{assump:key_sae1}), where a superpopulation model learned from $\mathcal{U}^s$ is used to predict the subpopulation-level target parameter based on the $N_j-n_j$ unknown values of $Y_{ij}$. These predictions are then used to compute the parameter of interest, as in \eqref{eq:tau_hat}. The predictive theory assumes that the superpopulation model provides sufficiently accurate predictions, with goodness of fit typically assessed by MSE relative to this model \citep{rao2015small, morales2021course}. The following assumption, along with Assumption~\ref{assump:key_sae1}, serves as a key SAE assumption without specifying the superpopulation model's form.
\begin{assumption}\label{a:key_sae2} Let 
$\hat{\mu}_n(\cdot)$ be an estimator of $\mu(\cdot)$ such that $\norm{\hat{\mu}_n(\cdot) - \mu(\cdot)} = O_p(n^{-1/2})$, where $\norm{f}^2$ denotes the $L^2$ norm of a function $f \in \mathcal{F}$, with $\mathcal{F}$ being a Donsker class.
\end{assumption}

Within the SAE framework, the estimator in \eqref{eq:tau_hat} can be obtained via two approaches: the \textit{global estimation strategy} (Section~\ref{sec:est_strat1}) and the \textit{local estimation strategy} (Section~\ref{sec:est_strat2}). In the global estimation strategy, all elements of the nuisance parameter $\bm{\eta}$ are estimated using data from all subpopulations, while subpopulation-specific features are accounted for through appropriate modeling (e.g., adding random effects, growing trees within each subpopulation). In contrast, the local estimation strategy learns $\mu(\cdot)$ from $n$ observations in $\mathcal{U}^s$ across all subpopulations and uses it to predict out-of-sample $\hat{Y}_{ij}$ for $i \in \mathcal{U}^r$ (see notation at the beginning of Section~\ref{sec:notation_est}). Then, the estimation of $\mu_a(\cdot)$ and $e_1(\cdot)$ is carried out locally at the subpopulation level using $N_j$ subpopulation-specific data.

\subsubsection{Global estimation strategy}\label{sec:est_strat1}
Under the global estimation strategy, we use available data 
to learn $\mu(\bm{X}_{A, ij})$, $\mu_a(\bm{X}_{ij})$ and $e_1(\bm{X}_{ij})$ which hold true for the entire population $\mathcal{U}$; the differences between subpopulations are accounted for by appropriate modelling and/or population-specific covariates. This strategy is summarised in Algorithm~\ref{al:ges}.
\begin{algorithm}
\caption{Global estimation strategy}\label{al:ges}
\begin{algorithmic}[1]
\State Estimate \( \mu(\bm{X}_{A, ij}) \) using \( \bm{Z}_{ij} \) for \( i \in \mathcal{U}^s_j \), and set \( \hat{Y}_{ij} = \hat{\mu}_n(\bm{X}_{A, ij}) \) for \( i \in \mathcal{U}^r_j \), and \( \hat{Y}_{ij} = Y_{ij} \) for \( i \in \mathcal{U}^s_j \), with \( j \in \mathcal{V} \).
\State Estimate $\mu_a(\bm{X}_{ij})$ using $\bm{Z}_{ij}$ for \( i \in \mathcal{U}^s_j \), $j\in \mathcal{V}$ and obtain $\hat{\mu}_{a,n}(\bm{X}_{ij})$ for  $i\in \mathcal{U}_j$, $j\in\mathcal{V}$. 
\State Estimate $e_1(\bm{X}_{ij})$ using $\bm{X}_{A, ij}$ for $i\in \mathcal{U}_j$, $j\in\mathcal{V}$, and obtain $\hat{e}_{1,N}(\bm{X}_{ij})$ for $i\in \mathcal{U}_j$, $j\in\mathcal{V}$. 
\State Use $\hat{Y}_{ij}$, $\hat{e}_{1, N}(\bm{X}_{ij})$ and $\hat{\mu}_{a, n}(\bm{X}_{ij})$ to construct estimators for $\tau_j$.
\end{algorithmic}
\end{algorithm}
\begin{example}\label{example:CSAE-OR} The CSAE outcome regression estimator (CSAE-OR) is defined as:  
\begin{equation*}\label{eq:CSAE-OR}
  \hat{\tau}_{or,j}= \hat{\tau}^1_{or,j}- \hat{\tau}^0_{or,j}=  
   \mathbb{P}_{N_j}   \{\hat{\mu}_{1, n}(\bm{X}_{ij})\}  -  \mathbb{P}_{N_j}   \{\hat{\mu}_{0, n}(\bm{X}_{ij})\}  .
\end{equation*} 
\end{example}
\begin{example}\label{example:CSAE-IPW}  The CSAE inverse probability weighted estimator (CSAE-IPW) is defined as: $$\hat{\tau}_{ipw,j} =
\hat{\tau}^1_{ipw,j} - 
\hat{\tau}^0_{ipw,j} =
\mathbb{P}_{N_j} \left\{ \frac{I(A_{ij} = 1)\hat{Y}_{ij}}{\hat{e}_{1,N}(\bm{X}_{ij})} \right\}- \mathbb{P}_{N_j} \left\{ \frac{I(A_{ij} = 0)\hat{Y}_{ij}}{\hat{e}_{0,N}(\bm{X}_{ij})} \right\}. $$
\end{example}
\begin{example}\label{example:CSAE-NIPW}  The CSAE normalised inverse probability weighting estimator (CSAE-NIPW) is defined as:
$$\hat{\tau}_{nipw,j} =  
\hat{\tau}^1_{nipw,j} - 
\hat{\tau}^0_{nipw,j} =
\mathbb{P}_{\hat{N}^1_j} \left\{ \frac{I(A_{ij} = 1)\hat{Y}_{ij}}{\hat{e}_{1,N}(\bm{X}_{ij})} \right\}
- \mathbb{P}_{\hat{N}^0_j} \left\{ \frac{I(A_{ij} = 0)\hat{Y}_{ij}}{\hat{e}_{0,N}(\bm{X}_{ij})} \right\}.
$$
\end{example}
\begin{example}\label{example:CSAE-AIPW} The CSAE augmented inverse probability weighted estimator (CSAE-AIPW)  is defined as:
\begin{align*}
\hat{\tau}_{aipw,j}  = \hat{\tau}^1_{aipw,j} - \hat{\tau}^0_{aipw,j} &= 
\mathbb{P}_{N_j} \left\{ \frac{I(A_{ij}=1)}{\hat{e}_{1,N}(\bm{X}_{ij})}\{\hat{Y}_{ij}-\hat{\mu}_{1,n}(\bm{X}_{ij})\}+\hat{\mu}_{1,n}(\bm{X}_{ij})\right\}  \\
&- 
\mathbb{P}_{N_j} \left\{ 
 \frac{I(A_{ij}=0)}{\hat{e}_{0,N}(\bm{X}_{ij})}\{\hat{Y}_{ij}-\hat{\mu}_{0,n}(\bm{X}_{ij})\}+\hat{\mu}_{0,n}(\bm{X}_{ij})\right\}.
\end{align*}
\end{example}

\subsubsection{Local estimation strategy}\label{sec:est_strat2}

Under the local estimation strategy, we first use the available data to learn $\mu(\bm{X}_{A, ij})$ and obtain predictions $\hat{Y}_{ij}$ for $i \in \mathcal{U}^r_j$, $j \in \mathcal{V}$. We then estimate the subpopulation-specific ATE using only subpopulation-specific data, with out-of-sample elements replaced by predictions \(\hat{\bm{Z}}^T_{ij} = (\hat{Y}_{ij}, \bm{X}_{ij}^T, A_{ij})^T \). Since imputation (or prediction) mitigates data scarcity, any ATE estimator/learner (e.g., AIPW, TMLE, DML) can be applied, cf. diagram in Figure~\ref{fig:EstimationStrategies} in Appendix~\ref{sec:SM_simulations} and Section~\ref{sec:sim_modelbased} in which we analyse their performance in sensitivity analysis. Although subpopulation-specific models are used, the local estimation strategy aligns with the predictive approach of the CSAE framework because data from the entire sample \(\mathcal{U}^s\) is pooled to predict missing \(Y_{ij}\) \(i \in \mathcal{U}^r\), $j\in \mathcal{V}$. Before detailing the steps of the local estimation strategy in Algorithm~\ref{al:les}, recall that $\bm{X}_{ij} = (\bm{W}^T_{ij}, G_{ij})^T$ and define \(\mu_{a, j}(\bm{W}_{ij}) \coloneqq E(\hat{Y}_{ij} \mid  \bm{W}_{ij}, A_{ij} = a, G_{ij} = j)\) and \(e_{1,j}(\bm{W}_{ij}) \coloneqq P(A_{ij} = 1 \mid \bm{W}_{ij}, G_{ij} = j)\). Furthermore, let \(\{\hat{\mu}_{a,N_j}(\cdot), \hat{e}_{1,N_j}(\cdot)\}\) be the estimators of \(\{\mu_{a,j}(\cdot), e_{1,j}(\cdot)\}\). 
\begin{algorithm}
\caption{Local estimation strategy}\label{al:les}
\begin{algorithmic}[1]
\State Estimate \( \mu(\bm{X}_{A, ij}) \) using \( \bm{Z}_{ij} \) for \( i \in \mathcal{U}^s_j \), and set \( \hat{Y}_{ij} = \hat{\mu}_n(\bm{X}_{A, ij}) \) for \( i \in \mathcal{U}^r_j \), and \( \hat{Y}_{ij} = Y_{ij} \) for \( i \in \mathcal{U}^s_j \), with \( j \in \mathcal{V} \).
\For{$j \in \mathcal{V}$}
\State Estimate \(\mu_{a, j}(\bm{W}_{ij}), e_{1,j}(\bm{W}_{ij}) \) using $\hat{\bm{Z}}_{ij}$ and obtain \(\hat{\mu}_{a, j}(\bm{W}_{ij}), \hat{e}_{1,j}(\bm{W}_{ij}) \) for $i\in \mathcal{U}_j$.
\EndFor
\State Use $\hat{Y}_{ij}$, $\hat{e}_{1,j}(\bm{W}_{ij})$ and $\hat{\mu}_{a,j}(\bm{W}_{ij})$ to construct estimators for $\tau_j$.
\end{algorithmic}
\end{algorithm}

We define four examples of CSAE methods following the local estimation strategy. 
\begin{example}\label{example:CSAE-OR-j} Subpopulation specific CSAE-OR estimator is defined as:
\begin{equation*}\label{eq:CSAE-OR-j}
  \hat{\tau}^L_{or,j}= \hat{\tau}^{L,1}_{or,j}- \hat{\tau}^{L,0}_{or,j}=  
   \mathbb{P}_{N_j}   \{\hat{\mu}_{1, N_j}(\bm{W}_{ij})\}  -  \mathbb{P}_{N_j}   \{\hat{\mu}_{0, N_j}(\bm{W}_{ij})\}. 
\end{equation*} 
\end{example}
\begin{example}\label{example:CSAE-IPW-j}  Subpopulation specific CSAE-IPW estimator is defined as: $$\hat{\tau}^L_{ipw,j} =
\hat{\tau}^{L,1}_{ipw,j} - 
\hat{\tau}^{L,0}_{ipw,j} =
\mathbb{P}_{N_j} \left\{ \frac{I(A_{ij} = 1)\hat{Y}_{ij}}{\hat{e}_{1,N_j}(\bm{W}_{ij})} \right\}- \mathbb{P}_{N_j} \left\{ \frac{I(A_{ij} = 0)\hat{Y}_{ij}}{\hat{e}_{0,N_j}(\bm{W}_{ij})} \right\}. $$
\end{example}
\begin{example}\label{example:CSAE-NIPW-j}  Subpopulation-specific CSAE-NIPW estimator~is defined as:
$$\hat{\tau}^L_{nipw,j} =  
\hat{\tau}^{L,1}_{nipw,j} - 
\hat{\tau}^{L,0}_{nipw,j} =
\mathbb{P}_{\hat{N}^{L,1}_j}^{N_j} \left\{ \frac{I(A_{ij} = 1)\hat{Y}_{ij}}{\hat{e}^L_{1,N_j}(\bm{W}_{ij})} \right\}
- \mathbb{P}_{\hat{N}^{L,0}_j}^{N_j} \left\{ \frac{I(A_{ij} = 0)\hat{Y}_{ij}}{\hat{e}^L_{0,N_j}(\bm{X}_{ij})} \right\}.
$$
\end{example}
\begin{example}\label{example:CSAE-AIPW-j}  Subpopulation-specific CSAE-AIPW estimator is defined as:
\begin{align*}
\hat{\tau}^L_{aipw,j}  = \hat{\tau}^{L,1}_{aipw,j} - \hat{\tau}^{L,0}_{aipw,j} &= 
\mathbb{P}_{N_j} \left\{ \frac{I(A_{ij}=1)}{\hat{e}_{1,N_j}(\bm{W}_{ij})}\{\hat{Y}_{ij}-\hat{\mu}_{1,N_j}(\bm{W}_{ij})\}+\hat{\mu}_{1,N_j}(\bm{W}_{ij})\right\}  \\
&- 
\mathbb{P}_{N_j} \left\{ 
 \frac{I(A_{ij}=0)}{\hat{e}_{0,N_j}(\bm{W}_{ij})}\{\hat{Y}_{ij}-\hat{\mu}_{0,N_j}(\bm{W}_{ij})\}+\hat{\mu}_{0,N_j}(\bm{W}_{ij})\right\}.
\end{align*}
\end{example}
Appendix~\ref{sec:asumpt_normality} contains Proposition~\ref{prop:normality} proving the normality of CSAE estimators, accompanied by a diagram illustrating global and local estimation strategies in Figure~\ref{fig:EstimationStrategies}. Both strategies rely heavily on nuisance parameters, which should be learned/estimated with the highest precision. Section~\ref{sec:description_nuisance_param} evaluates various methods for this purpose, including general approaches and those specifically designed for the SAE setting.

\subsection{Other estimators}\label{sec:les}
This section introduces two benchmark estimators used in our sensitivity analysis (Section~\ref{sec:simulations}) and the case study (Section~\ref{sec:data_analysis_impact}). In Example~\ref{example:did}, we define a direct estimator applied in the preliminary data analysis (Section~\ref{sec:impact_job}; see Figures~\ref{fig:map_comp}--\ref{fig:sate}), whereas in Example~\ref{example:d-ipw} we present a survey-specific IPW estimator.
\begin{example}\label{example:did} 
Let $\hat{N}^a_{H,j} = \sum_{i = 1}^{n_j^a}w_{ij}$, $w_{ij}=1/\pi_{ij}$ be a design (or basic) weight and $\pi_{ij}$ be a selection probability \citep{sarndal1992model}. Hajek-type estimator is defined as 
 $$\hat{\tau}_{H,j} = \hat{\tau}^1_{H,j} - \hat{\tau}^0_{H,j}  = \frac{1}{\hat{N}^1_{H,j}}\sum_{i = 1}^{n^1_j}
  w_{ij}Y_{ij}
- \frac{1}{\hat{N}^0_{H,j}}\sum_{i = 1}^{n^0_j} w_{ij} Y_{ij} .$$
\end{example}
\begin{example}\label{example:d-ipw} Survey-specific IPW estimator is defined as: 
$$\hat{\tau}_{dir,j} =
\hat{\tau}^1_{dir,j} - 
\hat{\tau}^0_{dir,j} =
\frac{1}{n_j}\sum_{i = 1}^{n_j} \left\{ \frac{I(A_{ij} = 1)Y_{ij}}{\hat{e}_{1,n}(\bm{X}_{ij})} \right\}- \frac{1}{\hat{N}^1_{H,j}}\sum_{i = 1}^{n^1_j}\left\{ \frac{I(A_{ij} = 0)Y_{ij}}{\hat{e}_{0,n}(\bm{X}_{ij})} \right\}. $$
\end{example}
However, as argued in Section~\ref{sec:impact_job}, estimators relying solely on sub-sample-specific data, such as those in Examples~\ref{example:did}–\ref{example:d-ipw}, are inefficient, as will become evident in Section~\ref{sec:sim_results}. Moreover, the estimator in Example~\ref{example:d-ipw} targets a different estimand of interest \citep[][]{stuart2011use, ranjbar2021causal}. Consequently, these estimators are not included in our CSAE framework.

\section{Estimation of variance of $\hat{\tau}_j$}\label{sec:boot_theory}

We introduce a novel and general bootstrap scheme specifically designed to construct percentile confidence intervals for \(\hat{\tau}_j\) in \eqref{eq:tau_hat}, offering a new approach to addressing this problem. The scheme combines elements of the random effects block bootstrap of \cite{chambers2013random} and the semiparametric random effects bootstrap of \cite{carpenter2003novel}. The success of our procedure hinges on sampling from appropriate sets of residuals and debiasing, the latter being crucial to achieving correct coverage of the confidence intervals for \(\tau_j\). Bias in the final estimates is a well-known issue in SAE, often arising from the introduction of the working model. 
It is typically addressed by providing mean squared errors of the estimates. Meanwhile, debiasing is also a standard practice in ML-based procedures \citep[cf.][]{chernozhukov2018, van2011targeted}.

Let us first define two sets of residuals. As  in Sections~\ref{sec:est_strat1}-\ref{sec:est_strat2}, let $\hat{\mu}_n(\bm{X}_{A, ij})$ be estimated value of $\mu(\bm{X}_{A, ij})$ obtained using  \( \bm{Z}_{ij} \) for \( i \in \mathcal{U}^s_j \) with \( j \in \mathcal{V} \). We define marginal residuals $r_{ij} = Y_{ij} - \hat{\mu}_n(\bm{X}_{A, ij})$ whereas level-1 residuals $r^{(1)}_{ij}$ and level-2 residuals $r^{(2)}_j$ are: 
\begin{align}\label{eq:residuals}
r_{ij}^{(1)} = r_{ij} - r^{(2)}_j,\; r^{(1)} = (r_{11}^{(1)}, \dots, r_{n_mm}^{(1)})^T, r^{(2)}_j = \sum_{i\in \mathcal{U}_j^s}^{} \frac{r_{ij}}{n_j} ,\; r^{(2)} = (r^{(2)}_1, \dots, r^{(2)}_m)^T,  
\end{align}
for $j\in \mathcal{V}, i\in \mathcal{U}_j^s$. We will focus on constructing bootstrap confidence intervals \( I_{1-\alpha} \) at the \( (1-\alpha) \)-level, whose asymptotic coverage probability converges to \( 1-\alpha \) as the number of bootstrap samples grows (\( B \to \infty \)) for \( \alpha \in (0, 1) \). 
Let $\hat{P}_j$ be an estimate of the distribution $P_j$, and let $\hat{\tau}^*_j$ be computed from observations generated according to $\hat{P}_j$. To construct intervals $I_{1-\alpha}$ using Efron's percentile method \citep{efron1981nonparametric}, it is sufficient to estimate appropriate quantiles from the distribution of $\hat{\tau}^*_j$ where the quantile at the level $\alpha$ is defined as $q_{j,\alpha}=\inf\{a\in \mathbb{R}: P_j(\hat{\tau}^*_j\leq a|\hat{P}_j)\geq \alpha\}$.  
While bootstrapping $B$ times, quantile $q_{j,\alpha}$ can be approximated using $q^*_{j,\alpha}$, the $[\{\alpha B\}+1]^{th}$ order statistics from the empirical distribution of the bootstrapped values $\hat{\tau}^*_j$. A detailed bootstrap scheme is presented in Algorithm~\ref{al:boot}.
\begin{algorithm}
\caption{Bootstrap scheme to obtain confidence intervals for $\hat{\tau}_j$}\label{al:boot}
\begin{algorithmic}[1]
\State Obtain \( \hat{\mu}_n(\bm{X}_{A, ij}) \) using \( \bm{Z}_{ij} \) for \( i \in \mathcal{U}^s_j \) with \( j \in \mathcal{V} \).
\For{$b = 1, \dots, B$}
\State Obtain vectors $r^{*(1, b)}\in \mathbb{R}^{n}$, $r^{*(2, b)}\in \mathbb{R}^{m}$ by sampling independently with replace-\hspace*{1.4em}ment from $r^{(1)}$ and $r^{(2)}$ given in \eqref{eq:residuals}.
\State Using $Y_{ij}^{*(b)} = \hat{\mu}_n(\bm{X}_{A,ij}) + r_i^{*(2)} + r_{ij}^{*(1)}$, simulate bootstrap sample data with $\bm{Z}^{*(b)}_{ij} = \hspace*{1.4em} (Y_{ij}^{*(b)}, \bm{X}_{ij}^T, A_{ij})^T$, $i \in \mathcal{U}_j^s, j\in \mathcal{V}$.
\State Obtain bootstrap  estimates $\hat{\tau}^{*(b)}_j$ using a local or a global estimation strategy. 
\EndFor
\State Obtain an estimate of bias $bias(\hat{\tau}_j^*) = 1/B\sum_{b=1}^{B}\hat{\tau}^{*(b)}_j -\hat{\tau}_j$ and de-biased bootstrap estimates  $ \hat{\tau}^{*(b)}_j - bias(\hat{\tau}_j^*)$, $j \in \mathcal{V}$, $b = 1, \dots, B$.
\State Estimate critical values $q^*_{\hat{\tau}_j,\alpha/2}$, $q^*_{\hat{\tau}_j,1-\alpha/2}$ by the $[\{(\alpha/2)B\}+1]^{th}$ 
 and $[\{(1-\alpha/2)B\}+1]^{th}$, respectively, order statistics of $\hat{\tau}^{*(b)}_j - bias(\hat{\tau}_j^*)$.
\end{algorithmic}
\end{algorithm}
Then, a percentile bootstrap interval for an general 
estimator $\hat{\tau}_j$ in \eqref{eq:tau_hat} is given by
\begin{equation}\label{eq:ind_interval}
  I_{j, 1-\alpha}: \{q^*_{\hat{\tau}_j,\alpha/2}, q^*_{\hat{\tau}_j,1-\alpha/2}\}, \quad j \in \mathcal{V}.
\end{equation}
For certain estimators, double-bootstrap bias correction improves accuracy, which held true in our case. We extended Algorithm~\ref{al:boot} with a double-bootstrap scheme, effectively reducing estimator bias (see Table~\ref{tab:boot_best}). Due to its similarity to Algorithm~\ref{al:boot}, we provide the details inAlgorithm~\ref{al:boot2} in Appendix~\ref{sec:double-boot}. The latter also includes a Lemma and a Corollary in which we prove the consistency of intervals in \eqref{eq:ind_interval}.


\section{Sensitivity analysis through simulations}\label{sec:simulations}
\subsection{Description of methods to estimate nuisance parameters}\label{sec:description_nuisance_param}

We assessed the performance of the estimators introduced in Section~\ref{sec:methodology} through simulations. For both local and global estimation strategies, the nuisance parameters \( \mu(\bm{X}_{A, ij}) \), \( \mu_a(\bm{X}_{ij}) \), and \( e_{1}(X_{ij}) \) were estimated using various methods, including linear and generalized linear models (L), median regression (M), M-quantile regression \citep[Mq;][]{chambers2006m}, random forests \citep[R;][]{athey2019generalized}, a tuned version (Rt), a clustered version (Rc), and a model with both tuning and clustering (Rct). Additional methods include hierarchical mixed effects models (H) \citep{rao2015small} and mixed random forests (Hf) \citep{krennmair2022flexible}, incorporating different random effects structures: (i) area-specific random intercepts (H1r, Hf1r), (ii) both random intercepts and treatment effects (H2r, Hf2r), and (iii) separate models for treated and non-treated units (H2m, Hf2m). We also employed gradient boosting (Gb) and its tuned version \citep[Gbt;][]{chen2016xgboost}, as well as the super learner algorithm \citep[S;][]{van2007super}, which combined all approaches. Under the local estimation strategy, we used the same methods for predicting \( \mu(\bm{X}_{A,ij}) \) but excluded mixed models, mixed random forests, M-quantile regression, and clustered random forests for \( \mu_{j,a}(X_{ij}) \) and \( e_{j,1}(X_{ij}) \), as these parameters were estimated locally without area effects. Table~\ref{tab:checked_models} in Appendix~\ref{sec:SM_simulations} summarizes the statistical and ML techniques used for nuisance parameter estimation.

Under the local estimation strategy, we also tested the performance of DML estimator and TMLE applied to each subpopulation after predicting the missing part of the population (see Algorithm \ref{al:les}). TMLE and DML are not directly applicable under global estimation strategy due to the scarcity of data at the level of subpopulations, see discussion in Section~\ref{sec:discussion}. We used existing R packages \texttt{mle3}, \texttt{mlr3learners}, \texttt{DoubleML}  \citep{DoubleML2020}, and \texttt{tmle3}, \texttt{sl3} \citep{coyle2021tmle3-rpkg}, which limited our choice of methods for estimating nuisance parameters to random forest, (G)LM, and boosting.  Due to their erratic behaviour, a well-known issue in both causal inference and survey sampling, we omitted the estimators $\tau_{ipw,j}$ in Example~\ref{example:CSAE-IPW} and $\tau^L_{ipw,j}$ in Example~\ref{example:CSAE-IPW-j}, all the other were obtained using the publicly available R package \texttt{causalSAE} \citep{causalSAE}. 
For each considered estimator, we tested the performance of all possible combinations of nuisance parameters. For example, we tested the performance of $208$ CSAE-NIPW in Example~\ref{example:CSAE-AIPW} (16 imputation methods times 12 ways to estimate propensity scores). To test the performance of our estimators, we used the mean squared error (MSE), where lower values are desirable, calculated over $K = 1000$ simulation runs and $m$ subpopulations. We also computed the percentage error of an estimator's MSE relative to the best-performing estimator $
\overline{\text{MSE}}_{\mathrm{best}} = \min_{l = 1, \dots, L} \overline{\text{MSE}}^l$, where $L$ is the total number of estimators tested:
\begin{equation*}\label{eq:ARB_RRMSE}
\overline{\text{MSE}} = \frac{1}{m}\sum_{j = 1}^m \text{MSE}(\hat{\tau}_j) = \frac{1}{m}\sum_{j = 1}^m \left\{\sum_{k=1}^K \frac{(\hat{\tau}_j^{(k)} - \tau_j)^2}{K}\right\}, \quad \mathrm{\%err} = \frac{\overline{\text{MSE}} - \overline{\text{MSE}}_{\mathrm{best}}}{\overline{\text{MSE}}_{\mathrm{best}}} \times 100.
\end{equation*}

\subsection{Simulation setup}\label{sec:sim_modelbased}
In this section, we construct a synthetic population. During its construction, we aimed to: (a) emulate the features of the data from our case study, and (b) ensure impartiality by avoiding favouritism toward any specific data generation method. In particular, we generate counterfactual outcomes from the following nonlinear models: $Y_{ij}(0)  = \log(c^0_j + X_{s,j} + \bm{X}^T_{ij}\beta_{01}  + \exp(\bm{X}^T_{ij}\beta_{02} ) + \varepsilon_{ij}(0))$,  $Y_{ij}(1)  =\log(c^1_j + X_{s,j}  + \bm{X}^T_{ij}\beta_{11}  + \exp(\bm{X}^T_{ij}\beta_{12} ) + \varepsilon_{ij}(1))$,
where $var\{Y_{ij}(0)\} = 1.135$,  $var\{Y_{ij}(1)\} = 1.219$, $X_{s,j}\sim \mathcal N (0, 2.6)$ is a  subpopulation-specific known covariates, and $c^0_j\sim Unif(1,2)$, $c^1_j\sim Unif(2,3)$ are subpopulation-specific unknown intercepts imitating the subpopulation-level heterogeneity which is not explained by the covariates. Here $\bm{X}_{ij}$ is a covariate vector generated from a 10-dimensional multivariate normal distribution with a mean of zero and a covariance matrix $\Sigma$, where $\Sigma_{kk} = 1$ and $\Sigma_{kl} = 0.5$ for $k \neq l$ ($k, l = 1, 2, \dots, 10$). The coefficients $\beta_{01}$, $\beta_{02}$, $\beta_{11}$, and $\beta_{12}$ are generated as follows: $\beta^{k}_{01j} \sim \mathcal{N}(0, 3)$, 
$\beta^{k}_{02j} \sim 0.1 \times \mathcal{N}(0, 3)$, $\beta^{k}_{11j} \sim \beta^{j}_{01j} + 2 \times \mathcal{N}(4, 3)$, $\beta^{k}_{12j} \sim \beta^{j}_{02j} + 0.1 \times \mathcal{N}(4, 3)$,
for $k = 1, 2, \dots, 10$ and $j = 1, 2, \dots, m$. This means that the coefficients are the same for units within each subpopulation but different across subpopulations. This approach mimics the clustering effect without explicitly relying on mixed modeling.

We generated a population with \( m = 41 \) subpopulations, each containing \( N_j = 1000 \) units, yielding a total population size of \( N = 41000 \). In each simulation run, the population remained fixed while only the sampled values changed. The true subpopulation-level treatment effects \( \tau_j \), shown in Figure~\ref{fig:estimates_best_AIPW_DML}, are given by  $\tau_j = \frac{1}{N_j}\sum_{i = 1}^{N_j} \{Y_{ij}(1) - Y_{ij}(0)\}$. All true values are positive, with a range of 1.97 -- slightly smaller than the naive estimator's range of 2.16 (Figure~\ref{fig:sate}). This intentional difference accounts for the naive estimator’s high variability, suggesting the true range in the case study is smaller. The treatment indicator is generated from a Bernoulli distribution, $A_{ij} \sim \text{Bernoulli}\{e_1(\bm{X}_{ij}, X_{s,j})\}$, $e_1(\bm{X}_{ij}, X_{s,j}) = \{\exp(\bm{X}^T_{ij}\alpha^j + X_{s,j}\alpha^j_{X_s})\}/\{1 + \exp(\bm{X}^T_{ij}\alpha^j + X_{s,j}\alpha^j_{X_{s}})\}$. The coefficients are drawn as follows: $\alpha^j_{k} \sim \mathcal{N}(2, 6)$ for $k = 1, \dots, 9$, $\alpha^j_{10} \sim \mathcal{N}(0, 0.25)$, and $\alpha^j_{Z} \sim \mathcal{N}(0, 0.25)$, with $j = 1, 2, \dots, m$. Finally, the range of treated observations (0.606 to 0.899) closely matches that in the case study (Table~\ref{tab:descriptive_stats}).


To match the case study setting (Table~\ref{tab:descriptive_stats}), we set the number of treated units as \( n_j^1 = 0.02 \times N_j^1 \). We then generated a frequency vector \( \bm{f}_{01} \) of length \( m \) as follows: 25 values from \( U(0.01, 0.5) \) (\( j = 1, \dots, 25 \)), 10 from \( U(0.51, 1) \) (\( k = 26, \dots, 35 \)), and 6 from \( U(0.9, 1) \) (\( l = 36, \dots, 45 \)). The number of controls was computed as $n_j^0 = \lceil f_{01,j} \times n^1_j \rceil$, yielding total sample sizes $n_j = n^1_j + n^0_j$. This resulted in \( n = 945 \), with a sampling fraction of \( f = 0.023 \), slightly higher than in the case study but not expected to affect method comparisons.



\subsection{Simulation results}\label{sec:sim_results}

\begin{figure}[ht]
\begin{floatrow}
\capbtabbox[0.35\textwidth]{%
\addtolength{\tabcolsep}{-0.3em}
\begin{tabular}{lllll}\hline
\multicolumn{5}{|c|}{CSAE-AIPW}  \\
\multicolumn{1}{|c}{} & &   & \multicolumn{2}{c|}{Cov (in \%)} \\
\multicolumn{1}{|c}{$\mu$} &  $e_1$ &  $ \mu_a$  & B           & \multicolumn{1}{c|}{DB}       \\\hline
\multicolumn{1}{|c}{H2r} & X   & M   & 91.40  & \multicolumn{1}{c|}{95.53}   \\
\multicolumn{1}{|c}{H2r} & S    & M  & 91.57   & \multicolumn{1}{c|}{94.82}   \\
\multicolumn{1}{|c}{H2r} & X   & Rc  & 89.12  & \multicolumn{1}{c|}{94.41}   \\
\multicolumn{1}{|c}{H2r} & X  & Rct & 89.52 & \multicolumn{1}{c|}{96.79}    \\      \hline    
 &  & &  & \\
\end{tabular}
}{%
\caption{Empirical coverage probabilities of bootstrap confidence intervals; B, bootstrap; DB, double-bootstrap.  \label{tab:boot_best}}
}%
\ffigbox[0.6\textwidth]{%
\includegraphics[width=0.65\textwidth]{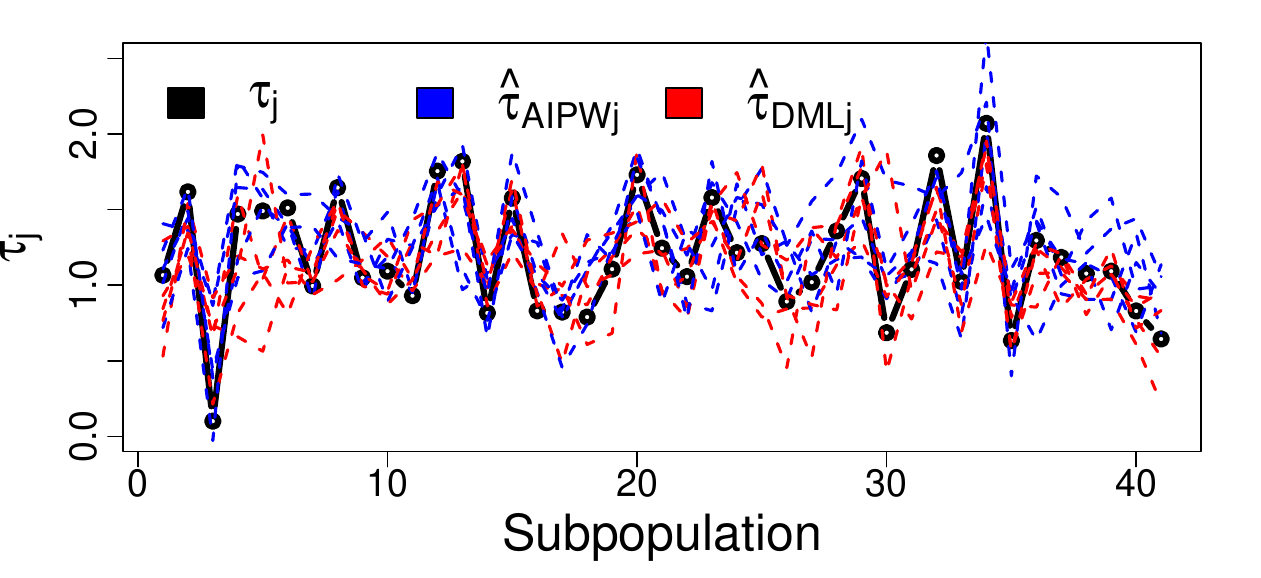}%
}{
\caption{Local DML (light blue) and global CSAE AIPW (light green) estimates of true ATE (dots) in simulations study.\label{fig:estimates_best_AIPW_DML}}%
}
\end{floatrow}
\end{figure}

Tables~\ref{tab:best_global}-\ref{tab:best_local_2} in Appendix~\ref{sec:SM_simulations} present the best-performing estimators under global and local estimation strategies. Among these, CSAE-AIPW estimators under global estimation strategy and DML estimators perform exceptionally well, see Table~\ref{tab:best_AIPW_DML}. The top estimators have MSE values no greater than 10\% above the best-performing estimator overall. This result is expected, as these estimators are theoretically guaranteed to achieve the lowest variance (i.e., they are efficient) when the imputation model closely approximates the true data-generating process. The estimator with the lowest MSE is the local DML estimator. However, as shown in Table~\ref{tab:best_AIPW_DML}, the computation time required for DML is significantly higher than for CSAE-AIPW, except in scenarios where the superlearner is used for propensity score estimation (a computationally intensive method). This demonstrates that CSAE-AIPW is scalable to large datasets, such as the one used in our case study, unlike DML (see the computational time for these two estimators in Section~\ref{sec:data_analysis_impact}).
\begin{table}[]
\begin{tabular}{|llllll|llllll|}\hline
\multicolumn{6}{|c|}{CSAE -AIPW}                          & \multicolumn{6}{c|}{DML}                                \\
$\mu$ & $e_1$ & $ \mu_a$ & MSE   & \% err & Time (s.) & $\mu$ & $e_1$ & $\mu_a$ & MSE   & \% err & Time (s.) \\\hline
H2r   & Gb     & M        & 0.107 & 5.659    & 2.989     & H2r   & Gb     & Gb       & 0.101 & 0.000    & 435.050  \\
H2r   & S     & M        & 0.108 & 7.061    & 1921.239  & H2m   & Gb     & Gb       & 0.105 & 4.265    & 435.1689  \\
H2r   & Gb     & Rc       & 0.109 & 7.793    & 10.965    & H2r   & L     & R       & 0.108 & 6.496    & 156.763  \\
H2r   & Gb     & Rct      & 0.109 & 7.902    & 13.012    & H2r   & L     & Gb       & 0.110 & 8.967    & 275.054  \\
H2r   & Gb     & R        & 0.109 & 8.264    & 9.771     & H2m   & L     & R       & 0.112 & 11.035   & 156.882 \\\hline
\end{tabular}
\caption{Best performing global CSAE-AIPW estimators and DML estimators. MSE, mean squared error; \% err, increase of MSE in percentage with respect to the best performing method; Time (s.), computational time.}
\label{tab:best_AIPW_DML}
\end{table}
Additionally, we are not aware of any method to accurately estimate the variability of DML after imputation. The pre-programmed variance estimation methods in the \texttt{DoubleML} package do not account for the additional variation introduced by imputation, leading to underestimated variance. However, the application of the best local DML estimator with $\hat{\eta}= \{\text{H2r}, \text{Gb}, \text{Gb}\}$ and the best global CSAE-AIPW estimator with $\hat{\eta}= \{\text{H2r}, \text{Gb}, \text{M}\}$ produces nearly identical point estimates, as shown in Figure~\ref{fig:estimates_best_AIPW_DML}. The figure also displays the true values of the ATE alongside estimates obtained from five randomly selected samples. Although the global CSAE-AIPW estimator exhibits slightly higher variability, the estimates are otherwise remarkably similar.

Finally, we evaluate the performance of Algorithm~\ref{al:boot} and its double bootstrap extension in Algorithm~\ref{al:boot2} for constructing well-performing confidence intervals. Our focus is on the class of computationally scalable and high-performing estimators, specifically the global CSAE-AIPW estimator (Table~\ref{tab:best_global}). The general simulation setup remains the same as in Section~\ref{sec:sim_modelbased}, except that in each simulation run, we generate $B = 1000$ bootstrap samples and compute confidence intervals $I_{1-\alpha, j}$ for $\tau_j$ as defined in \eqref{eq:ind_interval}. To assess the performance of our bootstrap scheme, we calculate the coverage of the intervals across simulation runs:
$\mathrm{Cov} =\frac{1}{mK}\sum_{j=1}^{m}\sum_{k=1}^{K}\bm{1} \{ \tau_j\in I^{k}_{j,1-\alpha} \}$.
As shown in Table~\ref{tab:boot_best}, the confidence intervals obtained using Algorithm~\ref{al:boot}are too narrow, with coverage probabilities below the desired 95\%. This issue is addressed by its double-bootstrap extension, which produces well-performing estimators regardless of the method used to estimate nuisance parameters.

\subsection{Discussion}\label{sec:discussion}
\textbf{Design-based versus model-based simulations}.
In this manuscript, we conducted a sensitivity analysis using a synthetic population designed to resemble the true population without assuming a specific model, ensuring an impartial estimator selection. However, one could argue that our results reflect the synthetic rather than the true population, and that design-based simulations, the gold standard in SAE, should have been used instead. Unfortunately, this was not feasible for two reasons. First, the small sample size in survey data, especially among treated units (see Table~\ref{tab:descriptive_stats}), prevented us from capturing the variability of true population samples. While we conducted hybrid simulations -- imputing missing out-of-sample values using parametric and machine learning methods -- these were ultimately model-based rather than true design-based simulations. Second, the lack of counterfactual outcomes made design-based simulations impractical. While \eqref{eq:iden_all} as in \citep[as in][]{ranjbar2021causal} could theoretically address this, the small sample size issue remains.
        
\textbf{Accounting for subpopulation-level variation in SAE}. The SAE literature \citep{rao2015small, morales2021course} emphasizes the need to account for subpopulation-level variation when predicting out-of-sample units for precise estimates -- an equally important principle for treatment effect estimation. Traditionally, such variation has been modeled using subpopulation-level random effects within frequentist and Bayesian frameworks. Building on this approach, some authors \citep{ krennmair2022flexible} have extended ML methods to incorporate random effects, applying them to SAE problems. In our manuscript, rather than modifying ML algorithms, we assume that the covariates include both individual-level and subpopulation-level information (see Section~\ref{sec:csae}), with the latter capturing subpopulation-level variability.

\textbf{Cross-fitting in causal SAE}.
Cross-fitting and sample-splitting are established methods to address ``double-dipping'' i.e., using the same data to estimate nuisance parameters \citep{van2011targeted, chernozhukov2018}. While we applied cross-fitting under the global estimation strategy, the small subpopulation sampling fractions and the need to predict out-of-sample outcomes significantly increased the variability of the final estimates. This made the results incomparable to those without cross-fitting or with cross-fitting applied post-prediction (see Table~\ref{tab:best_AIPW_DML}). As a result, we excluded them from the manuscript.


\section{Data analysis}\label{sec:data_analysis_impact}

In this section, we analyze how job stability influences relative poverty across 41 provinces in Italy. Before delving into the modeling details, we first examine the descriptive statistics of sample and subpopulation sizes, as shown in Table~\ref{tab:descriptive_stats}.  
\begin{table}[]
\begin{tabular}{|cccccccccc|}\hline
       & $N_j$   & $N_j^0$ & $N_j^1$ & $N_j^1/N_j$  & $n_j$& $n_j^0$ & $n_j^1$ & $n_j^1/n_j$ & $f_j$\\\hline
       min     & 17558  & 2658 & 14306 & 0.606  & 9 &1  & 8  & 0.800  & 0.00015  \\
       median    & 44989     & 8637 & 37522       & 0.834 & 64 &7  & 58 &0.904   & 0.00138\\
       mean    & 70103          & 12830   & 57273      & 0.803  & 106 & 11   & 95  &0.901  & 0.00176   \\
      max      & 328680 & 44723 & 292093  & 0.899 & 454 & 50   & 404 & 0.960& 0.00530 \\\hline            \end{tabular}
\caption{Descriptive statistics of sample and subpopulation sizes in our case study.}
\label{tab:descriptive_stats}
\end{table}
The sampling fraction at the subpopulation level is negligible, justifying the application of small area estimation principles. Additionally, the number of treated units (household heads with an open-ended contract) exceeds the number of control units (household heads with a temporary contract). This difference is particularly pronounced among the sampled units in each subpopulation, as explored further in the sensitivity analysis in Section~\ref{sec:sim_modelbased}.  

The outcome variable, $Y_{ij}$, is the log of equivalised household income, while the treatment variable, $A_{ij}$, represents the type of contract (1 for an open-ended contract and 0 for a temporary/fixed-term contract). Based on data availability and expert knowledge, we selected a set of covariates at the household level 
and the province level (see Table~\ref{table:variables} in the Appendix~\ref{sec:SM_data_applicaiton} for the full description of covariates).

\begin{figure}[ht]
    \centering
    \begin{subfigure}[t]{0.24\textwidth}
        \centering
 \includegraphics[width=\textwidth]{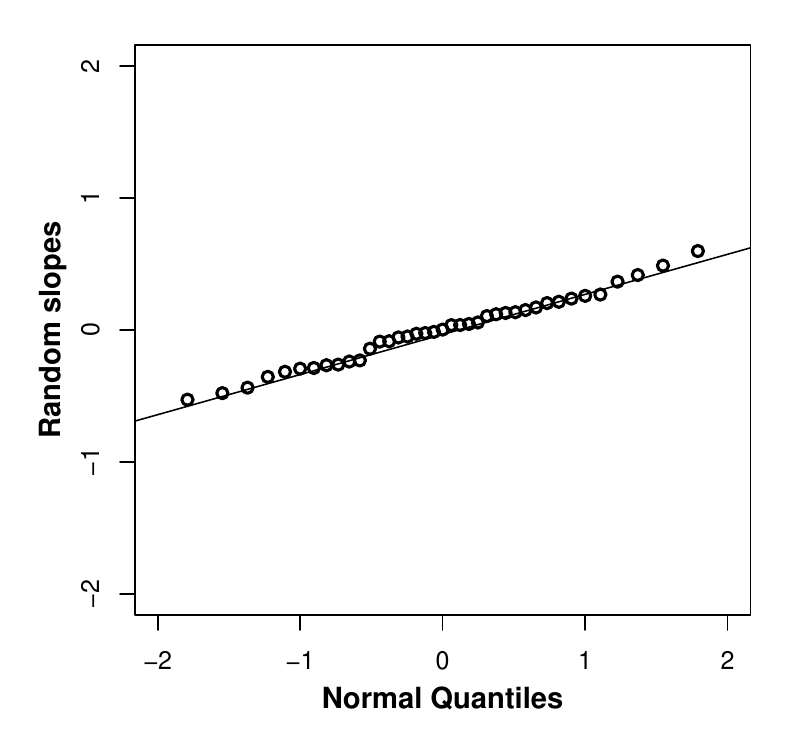}
        \caption{True data.}
        \label{fig:tau18}
    \end{subfigure}
    \begin{subfigure}[t]{0.24\textwidth}
        \centering
  \includegraphics[width=\textwidth]{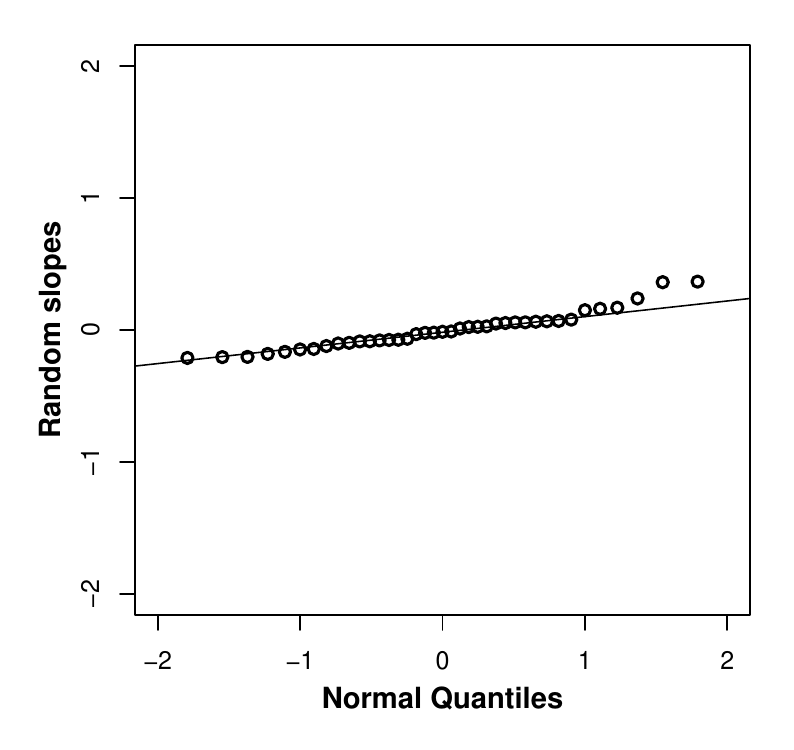}
        \caption{Simulataed data.}
        \label{fig:tau2}
    \end{subfigure}
       \begin{subfigure}[t]{0.24\textwidth}
        \centering
\includegraphics[width=\textwidth]{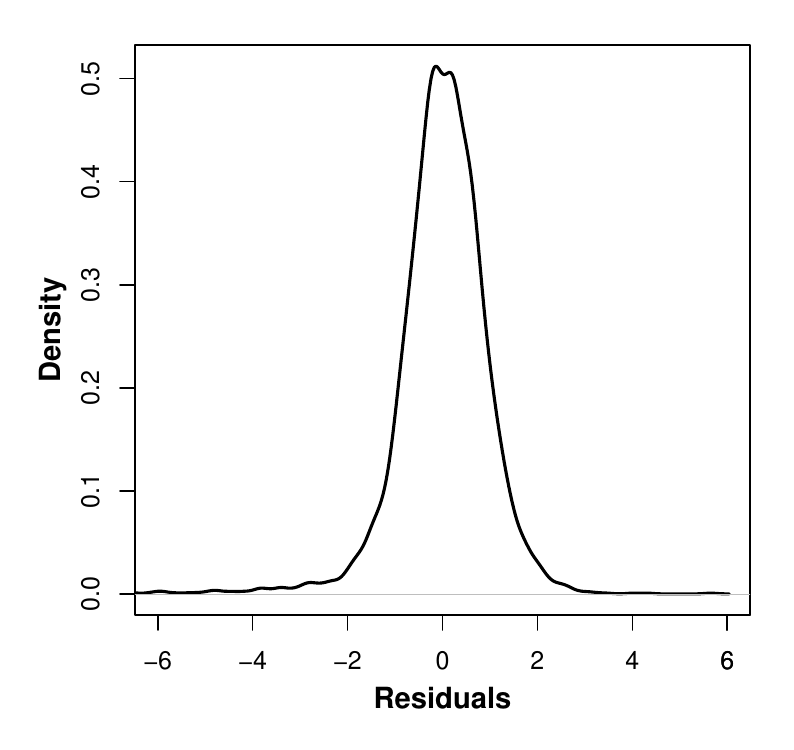}
        \caption{True data.}
        \label{fig:tau31}
    \end{subfigure}
    \hfill
    \begin{subfigure}[t]{0.24\textwidth}
        \centering
\includegraphics[width=\textwidth]{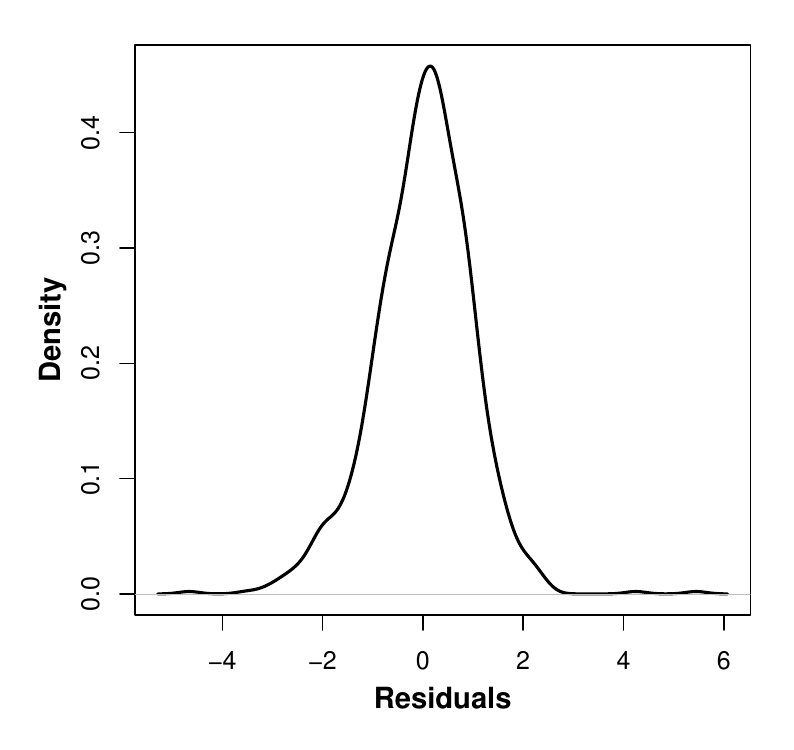}
        \caption{Simulated data.}
        \label{fig:tau22}
    \end{subfigure}
     \caption{QQ plots of random slopes and the density of residuals while fitting model to simulated and true data.}
    \label{fig:four_figures}
\end{figure}

As highlighted in Section~\ref{sec:simulations}, selecting an appropriate method for estimating \( \mu(\cdot) \) and predicting out-of-sample outcomes is critical. Based on Section~\ref{sec:sim_results}, we adopt a hierarchical model with area-specific random intercepts and treatment-specific random slopes, which outperformed others in our analysis and was effective in \cite{ranjbar2021causal}.  We also evaluated models with different covariate subsets (Table~\ref{table:variables}) using the Akaike Information Criterion (AIC), incorporating interactions between  covariates and treatment indicators as recommended by \cite{arpino2011specification,ranjbar2021causal}. AIC selected the model with all covariates and interactions between treatment variable \( A_{ij} \) and head of household-level features. Figure~\ref{fig:four_figures} displays QQ plots of random slopes and the density of residuals from the application of the model to both the case study and the simulations (results obtained from a single sample). As shown, there is a close similarity between the two, although the residuals in the data example exhibit greater skewness.

Based on the sensitivity analysis results in Section~\ref{sec:sim_results}, the local DML estimator with $\hat{\eta}= \{\hat{\mu}_{n}(\cdot), \hat{e}_{1, N}(\cdot), \hat{\mu}_{a,n}(\cdot)\}$ $ = \{\text{H2r}, \text{Gb}, \text{Gb}\}$ and the global CSAE-AIPW estimator with $\hat{\eta}= \{\hat{\mu}_{n}(\cdot), \hat{e}_{1, N}(\cdot), \hat{\mu}_{a,n}(\cdot)\}= \{\text{H2r}, \text{Gb}, \text{M}\}$ demonstrated the best performance across all estimation methods. Notably, applying these two estimation strategies produced nearly identical estimates across provinces in Italy (see Figure~\ref{fig:estimates_DML_AIPW_map} in Appendix~\ref{sec:SM_data_applicaiton}), confirming their close similarity in data analysis. However, reliable confidence intervals can only be constructed for the latter. Furthermore, the computational time required for the local DML estimator in our case study was prohibitive -- on an Intel(R) Core(TM) i9-9980HK CPU @ 2.40GHz with 32 GB RAM, obtaining the estimates took nearly three hours. This makes the use of computationally intensive methods for constructing confidence intervals infeasible. Therefore, in what follows, we focus on analysing global CSAE-AIPW estimates.

\begin{figure}[ht]
    \centering
     \hspace{-0.2cm}
    \begin{subfigure}[t]{0.6\textwidth}
        \centering
 \includegraphics[width=\textwidth]{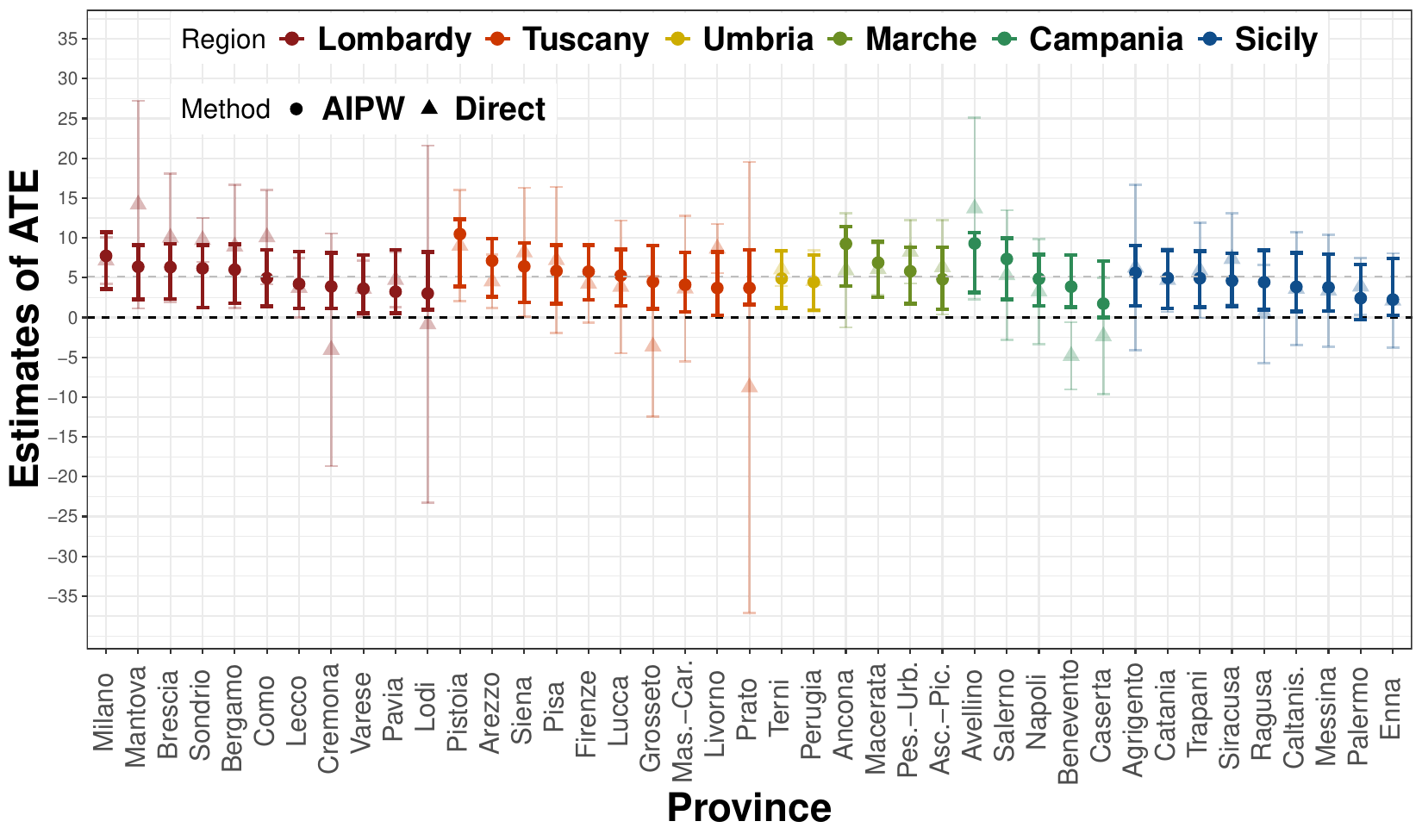}
        \label{fig:tau1}
    \end{subfigure}
    \hspace{-1cm}
    \begin{subfigure}[t]{0.37\textwidth}
        \centering
  \includegraphics[width=\textwidth]{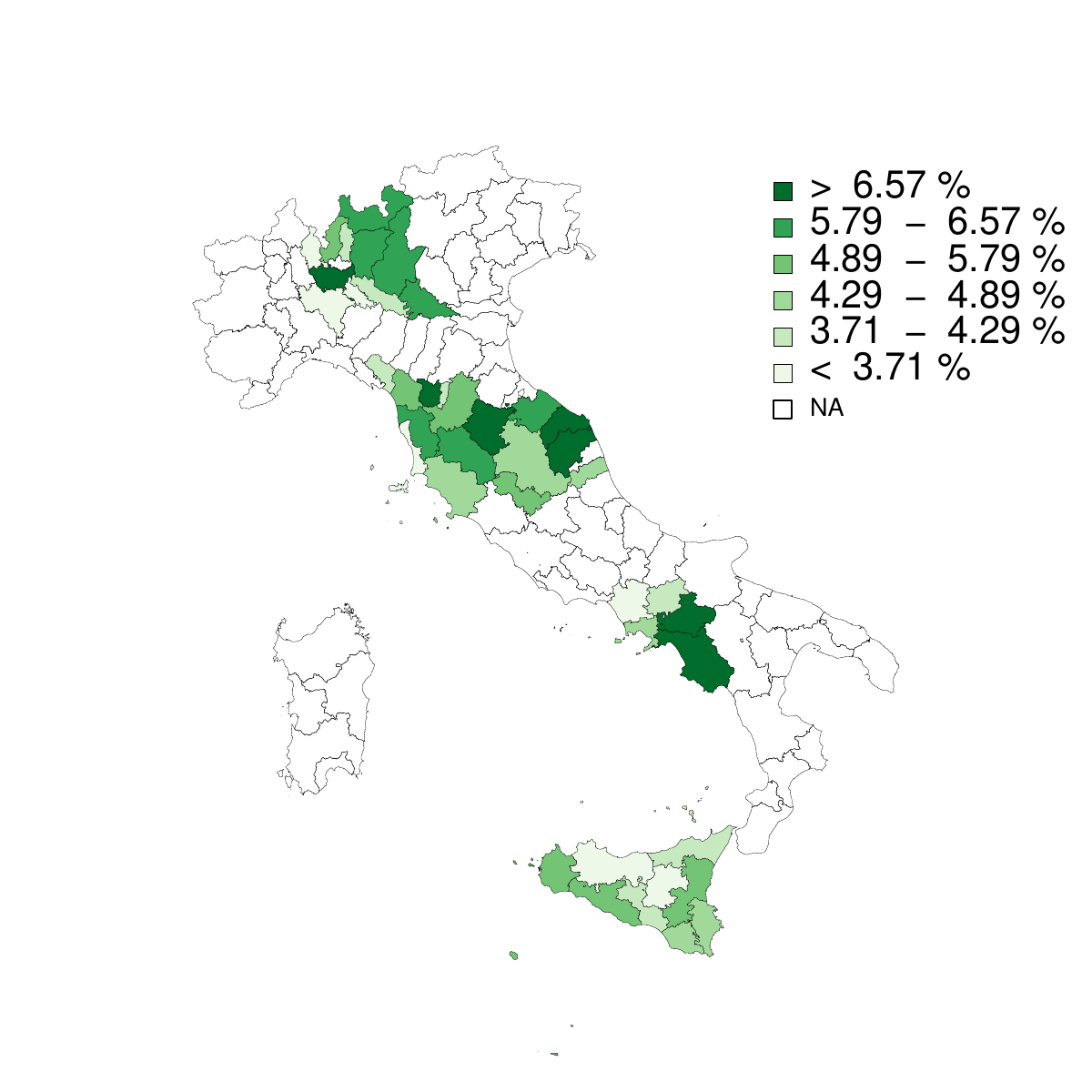}
        \label{fig:tau24}
    \end{subfigure}
     \caption{Point estimates of ATE (\%) with confidence intervals from global AIPW (darker) and direct estimators (lighter) (left panel). Map of global AIPW estimates (right panel).}
    \label{fig:estimates_AIPW_Directs}
\end{figure}

Figures~\ref{fig:estimates_AIPW_Directs} present the global CSAE-AIPW estimates from Example~\ref{example:CSAE-AIPW} alongside direct estimates from Example~\ref{example:did} across provinces in Italy's regions. Similarly to Figure~\ref{fig:map_comp}, we plot the ATE as percentages, specifically $\hat{\tau}_j/\hat{\tau}_j^{0} \times 100$, to facilitate interpretation. This representation directly shows the percentage increase in log-equivalised household income associated with holding a permanent job compared to a short-term contract. When comparing the CSAE-AIPW and direct estimates, it is evident that the confidence intervals for the latter are significantly wider than the bootstrap intervals for CSAE-AIPW. The mean confidence interval length is 14.55 for direct estimates versus 7.27 for CSAE-AIPW, making the former approximately twice as long. Additionally, some provinces exhibit negative direct estimates with very wide confidence intervals. In contrast, CSAE-AIPW estimates are all positive, apart from Palermo, where the lower boundary of the confidence interval crosses zero. These findings suggest that contract type influences household poverty levels, with permanent contracts providing greater income security and stability, leading to higher consumption and earnings. 

The differences between CSAE-AIPW estimates are less pronounced than those observed in the direct estimates (see also the map of estimates in Figure~\ref{fig:estimates_DML_AIPW_map}). Nevertheless, some regional differences are still apparent. The effect of job stability is less heterogenous and lower in Sicily (south) compared to Lombardy and Tuscany (north and central Italy). In fact, 4 of the 10 provinces with the lowest levels of job stability effects (Enna, Palermo, Messina, and Caltanissetta) are located in Sicily. Our results are in alignment with previous studies, cf. literature mentioned in Section \ref{sec:literature_review}. However, confidence intervals across all provinces intersect the grey dashed line, which represents the average effect of job stability ($5.18\%$) across all estimates.  In Campania (southern Italy), there is considerable heterogeneity among estimates, ranging from the second-highest value of $9.31\%$ in Avellino to the lowest value of $1.73\%$ in Caserta. In Lombardy (northern Italy), provinces with higher estimates of $\tau_j$ are concentrated in the south-west and northeast, while those with lower estimates are situated centrally.  In Umbria (central Italy), both point estimates of the effect of job stability are below the national average. In Tuscany and Marche (central Italy) the average values of estimates is above the national average. 


\section{Conclusions}\label{sec:discussion}
The primary objective of this study was to assess how contract type influences relative poverty across 41 Italian provinces. Motivated by this applied problem, we developed a comprehensive causal small area estimation (CSAE) framework for heterogeneous treatment effect estimation in settings where only a negligible fraction of outcomes is observed at the subpopulation level. To the best of our knowledge, this is the first study to propose a holistic approach to causal effect estimation in such a context. Through extensive sensitivity analyses, we demonstrated that the global AIPW-type estimator is computationally scalable and achieves one of the lowest MSEs among the considered methods, making it an ideal candidate for our case study. Additionally, we confirmed that the choice of predictive methods for out-of-sample units is critical, as it has the most significant impact on the performance of the estimation strategy. 

Our CSAE framework offers numerous avenues for further development. Future work could explore alternative methods for estimating nuisance parameters, such as multiple imputation \citep{rubin2004multiple, little2019statistical}, to better handle unobserved population data. The framework could also incorporate other estimators, like difference-in-differences, or refine machine learning techniques to address clustering and data scarcity at the subpopulation level. Beyond job stability, CSAE has broad applications, including health studies by the US National Center for Health Statistics, which employs synthetic estimation for subpopulation analysis \citep{rao2015small}. Similarly, the US National Agricultural Statistics Service estimates county-level crop acreage using satellite data and surveys. In this context, CSAE could assess the impact of interventions such as excessive fertilization across regions.

\appendixpage
\appendix
\renewcommand\thefigure{\thesection.\arabic{figure}}
\renewcommand\thetable{\thesection.\arabic{table}}

\section{Additional derivations and simulation resutls}\label{sec:appendix}
\subsection{Derivation of the identification formula}\label{appendix:iden}

Similary as in the main document, let $\hat{\bm{\eta}}= \{\hat{\mu}_{n}(\cdot), \hat{\mu}_{a,n}(\cdot), \hat{e}_{1, N}(\cdot)\}$ be some estimators of $\bm{\eta}= \{\mu(\cdot), \mu_{a}(\cdot), e_{1}(\cdot)\}$ which converge to potentially misspecififed limits $\bar{\bm{\eta}} \coloneqq \{\bar{\mu}(\cdot), \bar{\mu}_{a}(\cdot), \bar{e}_{1}(\cdot)\}$. In this section, we sketch the main arguments to establish the identification of the causal estimand $\tau_j(a)$ through the statistical parameter $\tau_j^a$. We start by proving that $E_j\{Y_{ij}(a)\} = E_j\{\mu_{a}(\bm{X}_{ij})\}$ using a g-computation formula \citep{hernan2020causal}:
\begin{align*}
E_j\{Y_{ij}(a)\} & 
=       E_j[E_j\{Y_{ij}(a)|\bm{X}_{ij}\} ] = 
E_j[E_j\{Y_{ij}(a)|\bm{X}_{ij}, S_{ij}= 1\} ] \\ &=   E_j[\{E_j\{Y_{ij}(a)|\bm{X}, A_{ij}= a, S_{ij} = 1\}] = E_j[\{E_j\{Y_{ij}|\bm{X}_j, A_{ij}= a, S_{ij} = 1\}]  ,\nonumber\\ & = E_j\{\mu_{a}(X_{ij})\}
\nonumber
\end{align*}
where the first equation follows by the law of total expectation, the second and the third by Assumption 2 in the main document, the fourth by causal consistency and the fifth by the definition of $\mu_{a}(\bm{X}_{ij})$. We have a similar identification formula using the propensity score weighting:
\begin{align}\label{eq:IPW_ident}
E_j\left\{\frac{I(A_{ij} = a)Y_{ij}}{e_a(\bm{X}_{ij})}\right\}
& = E_j\left[E_j\left\{\frac{I(A_{ij}=a)Y_{ij}}{e_a(\bm{X}_{ij})}\Big|\bm{X}_{ij}, S_{ij} = 1\right\}\right]\nonumber\\
&=E_j\left[\frac{Y_{ij}(a)}{e_a(\bm{X}_{ij})}E_j\left\{I(A_{ij}=a)|\bm{X}_{ij}, S_{ij} = 1\right\}\right] =  \{Y_{ij}(a)\}, 
\end{align}
where the first equality follows by the law of total expectation and Assumption~2 in the main document, the second by the causal consistency and the third by the definition of the propensity score in Assumption~3 in the main document. Finally, we can use both conditional expectations and propensity score weighting to identify $\tau_j(a)$ and use it to construct the augmented IPW (AIPW)-type estimators of $\tau^a_j$. If the propensity score model $e_1(\bm{X}_{ij})$ is correctly specified, we have for any $\bar{\mu}_{t}(\bm{X}_{ij})$ 
\begin{align*}
&E_j\left\{
\frac{I(A_{ij} = a)Y_{ij} }{e_a(\bm{X}_{ij})}- \frac{I(A_{ij} = a)-e_a(\bm{X}_{ij})}{e_a(\bm{X}_{ij})}\bar{\mu}_{a}(\bm{X}_{ij})\right\}  \\
&= E_j\{Y_{ij}(a)\} - E_j\left[\frac{E_j\left\{I(A_{ij}=a|\bm{X}_{ij}, S_{ij}=1)\right\}}{e_a(\bm{X}_{ij})}\bar{\mu}_a(\bm{X}_{ij})\right] +  E_j\{\bar{\mu}_a(\bm{X}_{ij})\} = \tau_j(a),
\end{align*}
where the second line follows by simple algebra, the identification formula in \eqref{eq:IPW_ident} and the law of total expectation. Similarly, if $\mu_{a}(\bm{X}_{ij})$ is correctly specified, then for any $\bar{e}_a(\bm{X}_{ij})$ we have 
\begin{align*}
&E_j\left\{
\frac{I(A_{ij} = a)Y_{ij} }{\bar{e}_a(\bm{X}_{ij})}- \frac{I(A_{ij} = a)-\bar{e}_a(\bm{X}_{ij})}{\bar{e}_a(\bm{X}_{ij})}\mu_{a}(\bm{X}_{ij})\right\}   \\	
&=    E_j\left(E_j\left[
\frac{I(A_{ij} = a)}{\bar{e}_a(\bm{X}_{ij})}\{Y_{ij}-\mu_a(\bm{X}_{ij})\} |\bm{X}_{ij},  S_{ij} = 1\right]\right) +  E_j\left\{\mu_{a}(\bm{X}_{ij})\right\} \nonumber\\
&=  E_j\left[
\frac{E_j\{I(A_{ij} = a)|\bm{X}_{ij},  S_{ij} = 1\}}{\bar{e}_a(X_{ij})}
E\left\{Y_{ij}(a)-\mu_{a}(\bm{X}_{ij}) |\bm{X}_{ij},  S_{ij} = 1\right\}\right] +  E\left\{\mu_{a}(\bm{X}_{ij})\right\}\\
&= E_j\left[	\frac{e_a(\bm{X}_{ij})}{\bar{e}_a(\bm{X}_{ij})}
\left\{E(Y_{ij}|\bm{X}_{ij},  S_{ij} = 1, A_{ij} = a)-\mu_{a}(\bm{X}_{ij}) \right\}\right] +  \tau_j(a)= \tau_j(a),
\end{align*}
where the first equality follows by the law of total expectation and by Assumption 2 in the main document, the second by (6) in the main document, and the third by the causal consistency, Assumption 2 in the main document, and the definition of $\mu_a(\bm{X}_{ij})$.

\subsection{Sketch of the asymptotic analysis of $\hat{\tau}_j$}\label{sec:asymp}
\subsection{Asymptotic normality of $\hat{\tau}_j$}\label{sec:asumpt_normality}
Given its superior performance in Section~6 in the main document, we focus on analyzing the asymptotic properties of $\hat{\tau}_{aipw,j}$ in Example~4 in the main document. The properties of the estimators presented in Examples~1-3 and Examples~5-8 in the main document can be derived similarly. Specifically, since the analysis of $\hat{\tau}_{aipw,j}$ reduces to studying $\hat{\tau}_{aipw,j}^a$, we focus on the latter without loss of generality. For simplicity, we streamline the notation by dropping the subscript, and throughout this section, we use $\hat{\tau}_j^a$ to represent $\hat{\tau}_{aipw,j}^a$.

Let  $\mathbb{P}_{N_j} = k^{-1} \sum_{i=1}^{N_j} \delta_{\bm{Z}_{ij}}$ denote the empirical distribution of the data, with $\delta_{\bm{Z}_{ij}}$ the Dirac measure. Throughout this section, we use $P_j\{f(\bm{Z}_{ij})\} = \int f(\bm{Z}_{ij}) d P_j $ to denote the expected value of $f(\bm{Z}_{ij})$ for a new observation $\bm{Z}_{ij}$, $i\in \mathcal{U}_j$, $j\in \mathcal{V}$. 
We further suppose that $\tau_j = P_j\{\varphi(\bm{Z}_{ij};\bm{\eta})\}$. Finally, let $\check{\bm{\eta}}= \{\mu(\cdot), \hat{\mu}_{a,n}(\cdot), \hat{e}_{1, N}(\cdot)\}$ which is a version of $\hat{\bm{\eta}}$ with $\hat{\bm{\mu}}(\cdot)$ replaced by $\mu(\cdot)$. To analyse the asymptotic behaviour of $\hat{\tau}_j^a- \tau_j^a$, we need to make two additional assumptions:


\begin{assumption}\label{a:conv_nuisance_param} 
Let $\hat{\bm{\eta}}  =\{\hat{\mu}_n(\cdot), \hat{\mu}_{a,K}(\cdot), \hat{e}_{1, K}(\cdot)\}$ be estimators that converge to \linebreak $\bar{\bm{\eta}}  =\{\bar{\mu}_n(\cdot), \bar{\mu}_{a,K}(\cdot), \bar{e}_{1, K}(\cdot)\}$ in a sense that $\norm{\hat{\bm{\eta}}(\cdot) - \bar{\bm{\eta}}(\cdot)}=o_p(1)$ where either (a) $\bar{\mu}_n(\cdot)=\mu(\cdot)$  and $\bar{\mu}_{a,n}(\cdot)=\mu_{a}(\cdot)$, or (b) $\bar{\mu}_n(\cdot)=\mu(\cdot)$,  and $\bar{e}_{1}(\cdot)=e_{1}(\cdot)$, but not necessarily both, hold true. \end{assumption}
\begin{assumption}\label{assump:conv_speed}
Let the sampling fractions $f = n/N$, $f_j = n_j/N_j$ be negligible in the following sense:
\begin{eqnarray*}
    \text{(i)}\lim_{n, N\to \infty}\frac{n}{N} = 0,\quad \text{(ii)}\lim_{n_j, N_j\to \infty}\frac{n_j}{N_j} = 0,\quad \text{(iii)} \lim_{n, N_j \to \infty}\frac{n}{N_j} = c, \text{ where } c \text{ is a finite constant. }
\end{eqnarray*}
\end{assumption}
Assumption~\ref{a:conv_nuisance_param} aligns with the double robustness property in classical causal inference, assuming the imputation model is correct. However, when $\bm{X}_{A,ij}$ is observed for the entire population and fully determines the propensity score, it seems more reasonable to assume version (b) of Assumption~\ref{a:conv_nuisance_param}. 
Finally, Assumption \ref{assump:conv_speed} describes the relationship between different quantities growing to infinity which will guarantee the convergence of $\hat{\tau}_j$. 

Proposition~\ref{prop:normality} shows that $\hat{\tau}_j^a$ is asymptotically normally distributed which allows us to construct confidence intervals based on the asymptotic theory only if we can find a way to estimate $\sigma^2_{\hat{\tau_j}}$ for which we use bootstrap (see Section~\ref{sec:est_variance} and Section~5 in the main document). 

\begin{proposition}\label{prop:normality}
    Let $\tau_j$ as defined in (7) in the main document, $\hat{\tau}_j^a$ as defined above, and $\sigma^2_{\hat{\tau}_j}$ the asymptotic variance of $\hat{\tau}_j^a$. If Assumptions 1-4 in the main document and Assumptions~\ref{a:conv_nuisance_param}-\ref{assump:conv_speed} are satisfied, then we have in distribution:
    \begin{equation}
        \sqrt{N_j}(\hat{\tau}^a_j-\tau^a_j ) \xrightarrow[]{d} \mathcal{N}(0,\sigma^2_{\hat{\tau}_j}).
    \end{equation}
\end{proposition}
\begin{proof}[Sketch of the proof.]
    
By extending the arguments of \cite{Kennedy2016, kennedy2023semiparametric} to our setting, we obtain the following decomposition
\begin{align}\label{eq:hat_tau_true_tau}
     \hat{\tau}_j^a- \tau_j^a  &=  \mathbb{P}_{N_j}\{\varphi(\hat{\bm{Z}}_{ij};\hat{\bm{\eta}})\}  -  P_j\{\varphi(\bm{Z}_{ij};\bm{\eta})\} \nonumber \\&= \mathbb{P}_{N_j}\{\varphi(\hat{\bm{Z}}_{ij};\hat{\bm{\eta}})\}  - \mathbb{P}_{N_j}\{\varphi(\bm{Z}_{ij};\check{\bm{\eta}})\} +  \mathbb{P}_{N_j}\{\varphi(\bm{Z}_{ij};\check{\bm{\eta}})\} +  P_j\{\varphi(\bm{Z}_{ij};\bm{\eta})\} \nonumber
     \\&=
     \mathbb{P}_{N_j}\{\varphi(\hat{\bm{Z}}_{ij};\hat{\bm{\eta}}) - \varphi(\bm{Z}_{ij};\check{\bm{\eta}})\}   +  (\mathbb{P}_{N_j} -  P_j)\varphi(\bm{Z}_{ij};\check{\bm{\eta}})+ 
     P_j
     \{\varphi(\bm{Z}_{ij};\check{\bm{\eta}}) - \varphi(\bm{Z}_{ij};\bm{\eta})\} \nonumber \\&= T_1 + T_2 + T_3,
\end{align}
where the first line follows by the definition, the second by adding and subtracting $ \mathbb{P}_{N_j}\{\varphi(\bm{Z}_{ij};\check{\bm{\eta}})\}$, and the third by adding and subtracting $ P_{j}\{\varphi(\bm{Z}_{ij};\check{\bm{\eta}})\}$. Next, for $T_1$ in  \eqref{eq:hat_tau_true_tau} we have: 
\begin{equation*}
\sqrt{N_j} T_1 =  \sqrt{N_j}\frac{1}{N_j}\sum_{i = 1}^{N_j-n_j}\left \{\left( \hat{Y}_{ij} - Y_{ij}  \right)  \frac{I(A_{ij} = a)}{\hat{e}_{a,N}(\bm{X}_{ij})}     \right\} \xrightarrow[]{d} \mathcal{N}(0, \sigma^2_{T_1}),
\end{equation*}
where the results follow by the Lindeberg-Feller central limit theorem if Assumptions~4 in the main document and \ref{assump:conv_speed} are correct. The asymptotics of terms $T_2$ and $T_3$ have been extensively studied in the literature \citep{pollard2012convergence,van2000asymptotic,van2011targeted,kosorok2008introduction}, as they commonly arise in the decomposition of classical AIPW estimators. Consequently, we only provide a brief discussion of these terms here. Specifically, $T_2$ is often called the empirical process term which equals $(\mathbb{P}_{N_j} -  P_j)\varphi(\bm{Z}_{ij};\bm{\eta}) + o_p(1/\sqrt{N_j})$ if Assumption \ref{a:conv_nuisance_param} holds and 
$\varphi(\bm{Z}_{ij};\check{\bm{\eta}})$ and $\varphi(\bm{Z}_{ij};\bm{\eta})$ belong to a Donsker class (or we use a sample splitting/cross fitting, see Remark~\ref{remark:splitting} and Section 6.4 in the main document). In addition, $T_3 = (\mathbb{P}_{N_j} - P_j)\psi(\bm{Z}_{ij};\bm{\eta}) + o_p(1/\sqrt{N_j})$ under Assumptions~3 in the main document and Assumption~\ref{a:conv_nuisance_param}, also following classical arguments in above mentioned literature. Following the arguments above and \cite{Kennedy2016}, $ T_2 + T_3$ is regular and asymptotically linear in a following sense 
$ T_2 + T_3 = (\mathbb{P}_{N_j} - P_j)\{\varphi(\bm{Z}_{ij};\bm{\eta}) + \psi(\bm{Z}_{ij};\eta)\} + o_p(1/\sqrt{N_j}) $, which means, by the theory of the semiparamteric inference \citep{van2000asymptotic} that $\sqrt{N_j}(T_2 + T_3) \xrightarrow[]{d} \mathcal{N}(0, \sigma^2_{T_2 + T_3})$
where $ \sigma^2_{T_2 + T_3}  =\mathbb{E}[\{\varphi(\bm{Z}_{ij};\bm{\eta}) + \psi(\bm{Z}_{ij};\bm{\eta})\}^2]$ which implies that $\sqrt{N_j}(\hat{\tau}^a_j-\tau^a_j ) \xrightarrow[]{d} \mathcal{N}(0,\sigma^2_{\hat{\tau}_j})$.
\end{proof}

\begin{remark}[Cross-fitting in causal SAE] \label{remark:splitting}
The sketch of asymptotic normality of $\hat{\tau}_{aipw,j}$ assumes that $\varphi(\bm{Z}_{ij};\bm{\check{\eta}})$ and $\varphi(\bm{Z}_{ij};\bm{\eta})$ belong to a Donsker class. This classical approach addresses ``double-dipping''—using the same data to estimate nuisance parameters $\bm{\eta}$ and the bias term $\mathbb{P}_{N_j}^{} {\varphi^{1}(\hat{\bm{Z}}_{ij}; \hat{\bm{\eta}})}$ -- which can lead to overfitting \citep[see, e.g., Figure 2 in][]{chernozhukov2018}. While well-established \citep{van2000asymptotic, kern2016assessing}, the Donsker assumption is restrictive, particularly in high-dimensional settings or with flexible methods for estimating nuisance parameters \citep{chernozhukov2018, kennedy2023semiparametric}. An alternative is to use sample-splitting or cross-fitting, which has been shown to perform well in various contexts \citep{kennedy_sharp, kennedy2023semiparametric, chernozhukov2018}, as demonstrated in Table~\ref{tab:best_local_2}, where cross-fitting is employed in the DML framework following out-of-sample outcome predictions. Nevertheless, in our setting, the application of cross-fitting before the imputation of the missing part of the subpopulation led to the substation increase of variability of the final estimator (see Section 6.4 for further discussion). Therefore, we did not pursue this path further. 
\end{remark}

\begin{remark}[Lack of double-robustness in a classical sense]
    In our setting, the double-robustness property of all estimators is replaced by a weaker condition, as outlined in Assumption~\ref{a:conv_nuisance_param}, which depends on the correctness of the imputation model. Two recent SAE manuscripts \citep{ranjbar2021causal, Schirripa2024} introduced IPW and NIPW estimators (Examples~2--3 in the main document) as ``double-robust'' in the classical sense. However, their arguments assume access to full population data, where these estimators coincide with the classical AIPW and satisfy double-robustness. In contrast, our empirical setting lacks access to out-of-sample outcomes, rendering their assumption unrealistic and necessitating a new class of estimators.
\end{remark}

\subsection{Estimation of variance of $\hat{\tau}_j$}\label{sec:est_variance}
\subsubsection{Double-bootstrap algorithm}\label{sec:double-boot}
Let $\hat{\mu}^*_n(\bm{X}_{A, ij})$ be a bootstrap estimate of $\hat{\mu}_n(\bm{X}_{A, ij})$ obtained using $\bm{Z}^{*(b)}_{ij} =  (Y_{ij}^{*(b)}, \bm{X}_{ij}^T, A_{ij})^T$, $i \in \mathcal{U}_j^s, j\in \mathcal{V}$ in Step 4 of Algorithm~3 in the main document. 
Bootstrap versions of  marginal residuals are defined as
$r^*_{ij} = Y^*_{ij} - \hat{\mu}^*_n(\bm{X}_{A, ij})$ whereas bootstrap level-1 residuals $r^{*(1)}_{ij}$ and bootstrap level-2 residuals $r^{*(2)}_j$ are given as follows
\begin{align}\label{eq:residuals2}
\hspace{-0.5cm}r^{*(2)}_j &= \sum_{i\in \mathcal{U}_j^s}^{} \frac{r^*_{ij}}{n_j},\; r^{*(2)} = (r^{*(2)}_1, \dots, r^{*(2)}_m)^T,r_{ij}^{*(1)} = r^*_{ij} - r^{*(2)}_j,\; r^{*(1)} = (r_{11}^{*(1)}, \dots, r_{n_mm}^{*(1)})^T, 
\end{align} 
for $j\in \mathcal{V}, i\in \mathcal{U}_j^s$. To obtain estimates of critical values using a double bootstrap scheme, we need to replace Steps 6-8 from of Algorithm~3 in the main document bt Steps 6-14 from Algorithm~\ref{al:boot2}.
\begin{algorithm}
\caption{Double bootstrap scheme to obtain confidence intervals for $\hat{\tau}_j$}\label{al:boot2}
\begin{algorithmic}[1]
\setcounter{ALG@line}{5} 
\State  Obtain \( \hat{\mu}^*_n(\bm{X}_{A, ij}) \) using \( \bm{Z}^*_{ij} \) for \( i \in \mathcal{U}^s_j \) with \( j \in \mathcal{V} \).
\For{$c = 1, \dots, C$}
\State Obtain vectors $r^{**(1, b)}\in \mathbb{R}^{n}$, $r^{**(2, b)}\in \mathbb{R}^{m}$ by sampling independently with replace-\hspace*{1.4em}ment from $r^{*(1)}$  and $r^{*(2)}$ given in \eqref{eq:residuals2}.
\State Using $Y_{ij}^{**(c)} = \hat{\mu}^*_n(\bm{X}_{ij}) + r_i^{**(2)} + r_{ij}^{**(1)}$, simulate bootstrap sample data with \hspace*{1.4em}$\bm{Z}^{**(c)}_{ij} = (Y_{ij}^{**(c)}, \bm{X}_{ij}^T, A_{ij})^T$, $i \in \mathcal{U}_j^s, j\in \mathcal{V}$.
\State Obtain double-bootstrap estimates $\hat{\tau}^{**(b)}_j$ using a local or a global estimation strategy. 
\EndFor
\State Obtain a bias corrected estimate of bias $bias^c(\hat{\tau}_j) = 2 bias(\hat{\tau}_j) - bias^*(\hat{\tau}_j^*) $ where $bias^*(\hat{\tau}_j^*) = 1/B\sum_{b=1}^{B}\hat{\tau}^{**(b)}_j -\hat{\tau}^*_j$.
\State De-biased bootstrap estimates  $\hat{\tau}^{*(b)}_j -bias^c(\hat{\tau}_j^*)$, $j \in \mathcal{V}, b = 1, \dots, B$.
\State Estimate critical values $q^*_{\hat{\tau}_j,\alpha/2}$, $q^*_{\hat{\tau}_j,1-\alpha/2}$ by the $[\{(\alpha/2)B\}+1]^{th}$ 
 and $[\{(1-\alpha/2)B\}+1]^{th}$, respectively, order statistics of $\hat{\tau}^{*(b)}_j - Bi^c(\hat{\tau}_j^*)$.
\end{algorithmic}
\end{algorithm}
We tested the performance of Algorithms~3 in the main document and Algorithm-\ref{al:boot2} for our best performing estimators in Section~6.3 in the main document. 

\subsubsection{Consistency of bootstrap confidence intervals}\label{sec:consistency}
To prove the consistency of $I_{j, 1-\alpha}$, we first note that quantiles obtained using Efron’s percentile method can be re-expressed in terms of the quantiles of a difference $\hat{\tau}^*_j-\hat{\tau}_j$. 
Without loss of generality, we choose $\widehat{\sigma}_{\hat{\tau}_j}$ to be independent of data, that is $\widehat{\sigma}_{\hat{\tau}_j} = \widehat{\sigma}_{\hat{\tau}^*_j} = 1$ \citep[cf. Section 21.1 in][]{van2000asymptotic}. Then, we have that $q_{j, \alpha} = \hat{\tau}_j + \xi_{j,\alpha}$ where $\xi_{j,\alpha}=\inf\{a\in \mathbb{R}: P(\hat{\tau}^*_j - \hat{\tau}_j\leq a|\hat{P}_j) \geq \alpha\}$. We exploit this connection to prove the asymptotic consistency of $I_{j, 1-\alpha}$. The first step is to prove Lemma~\ref{lemma:consistency_t_boot} which says that cumulative distribution functions of $\hat{\tau}^*_j - \hat{\tau}_j$ and $\hat{\tau}_j - \tau_j$ converge to the same limit. 

\begin{lemma}[Consistency of  $\tau^*_j - \hat{\tau}_j$]\label{lemma:consistency_t_boot} Let $\widehat{\sigma}_{\hat{\tau}_j} = \widehat{\sigma}_{\hat{\tau}^*_j} = 1$. If Assumptions~1-4 in the main document and 
Assumption~\ref{assump:conv_speed} hold, and the number of bootstrap samples grows ($B\to \infty$) then one has  in probability
    \begin{equation*}
    \sup_{a\in \mathbb{R}}\left\lvert P\left(\frac{\hat{\tau}_j- \tau_j}{\widehat{\sigma}_{\hat{\tau}_j}}\leq a\right)-
P\left(\frac{\hat{\tau}^*_j- \hat{\tau}_j}{\widehat{\sigma}_{\hat{\tau}^*_j}}\leq a|\hat{P}_j\right)\right\rvert\to 0.
    \end{equation*}
\end{lemma}
\begin{proof}
Let $\widehat{\sigma}_{\hat{\tau}_j} = \widehat{\sigma}_{\hat{\tau}^*_j} = 1$. Then we argue that, for every \(a\), \(F_{N_j}(a) = P(\hat{\tau}_j- \tau_j\leq a) \to P(T_j\leq a) = F_j(a)\) in distribution and \(F^*_{N_j}(a) = P(\hat{\tau}^*_j- \tau^*_j\leq a|\hat{P}_j) \to P(T_j\leq a) = F_j(a)\) in probability,  given the original sample size, where $T_j$ is a random variable with a continuous distribution function \(F_j(a)\). Without the loss of generality, we assume that $\hat{\bm{\vartheta}} = (\hat{\bm{\tau}}_j, \hat{\bm{\eta}}^T)^T$ is a solution to some estimating equation 
$\mathbb{P}_{N_j}\{m(\hat{\bm{Z}}_{ij};\hat{\bm{\vartheta}})\} = 0$, whereas $\hat{\bm{\vartheta}}^* = (\hat{\bm{\tau}}^*_j, \hat{\bm{\eta}}^{*T})^T$ is a solution to a bootstrap version of the same equation $\mathbb{P}_{N_j}\{m(\hat{\bm{Z}}^*_{ij};\hat{\bm{\vartheta}}^*)\} = 0$. Let $E^*$ be a bootstrap operator of the expected value. Then, under Assumption~\ref{assump:conv_speed} and Assumption~2 in the main document, it follows that $E^*[P^*\{m(\hat{\bm{Z}}^*_{ij};\hat{\bm{\vartheta}}^*  )\} ]= 0$ at $\hat{\bm{\vartheta}}^* = \hat{\bm{\vartheta}}$ which yields the consistency of the sequence of bootstrap estimators $\hat{\vartheta}^*$. One thus have that $\sqrt N_j(\hat{\tau}^*_j- \tau^*_j)$ and $\sqrt N_j(\hat{\tau}_j- \tau_j)$ converge to the same limiting distribution. 
\end{proof}
Corollary \ref{cor:intervals} ensures the consistency of intervals $I_{j, 1-\alpha}$ introduced in (10) in the main document. 
\begin{corollary}[Consistency of $I_{j,1-\alpha}$] \label{cor:intervals}
Lemma \ref{lemma:consistency_t_boot} implies that under the same assumptions one has
\begin{equation*}
P(\tau_j \in I_{j,1-\alpha}) \to 1-\alpha, \quad \alpha\in (0,1).
\end{equation*}
\end{corollary}

\begin{proof}
    The proof follows along the same lines as the second part of Lemma 23.3 of \cite{van2000asymptotic}. By Lemma~1, the sequence of distribution functions $F_{N_j}(a)$ 
converges weakly to $F_j$ which implies that corresponding quantile functions $F^{-1}_{N_j}(\alpha)$ converge to $F^{-1}_{j}(\alpha)$ at every continuity point, $\alpha\in (0,1)$. Similarly, one can conclude that quantiles $\xi_{j,\alpha}=F^{*-1}_{N_j}(\alpha) $ obtained from random distribution functions \(F^*_{j}(a) \) 
converge to $F^{-1}(\alpha)$ for every continurity point, but this time almost surely. Then, by Slutsky's lemma, $\hat{\tau}^*_j- \tau^*_j-\xi_{j,\alpha}$ converges to $T-F^{-1}(\alpha)$ and it follows that:
\begin{equation}\label{eq:quant}
P\left(\tau_j\geq \hat{\tau}_j - \xi_{j,1-\alpha}\right) = P\left( \hat{\tau}_j- \tau_j \leq  \xi^*_{j,1-\alpha} \right)\to P\left\{ T\leq F^{-1}(\alpha)\right\} = \alpha.
\end{equation}
Recall that $q_{j, \alpha} = \hat{\tau}_j + \xi_{j,\alpha}$ where $\xi_{j,\alpha}=\inf\{a\in \mathbb{R}: P(\hat{\tau}^*_j - \hat{\tau}_j\leq a|\hat{P}_j) \geq \alpha\}$ and assume that $F^*_{j}(a)$ is symmetric around zero. Then, we can rewrite equation in \eqref{eq:quant} as follows
\begin{equation}\label{eq:quant2}
P\left(\tau_j\geq  q_{j,1-\alpha}\right) = P\left( \hat{\tau}_j- \tau_j \leq  -\xi^*_{j,1-\alpha} \right)\to P\left\{ T\leq - F^{-1}(\alpha)\right\} = 1-\alpha.
\end{equation}
The final consistency of the intervals $I_{j, 1-\alpha}$ follows by assuming that $B\to \infty$.
\end{proof}

\section{Additional results from sensitivity analysis}\label{sec:SM_simulations}
Table~\ref{tab:checked_models} summarises statistical and machine-learning techniques used to obtain nuisance parameters for each estimator, whereas Figure~\ref{fig:EstimationStrategies} presents the diagram with local and global estimation strategies. 
\begin{table}[]
\begin{tabular}{|ll|llllll|lllll|lllll|}\hline
       &  & \multicolumn{6}{c|}{Global}                                                          & \multicolumn{10}{c|}{Local}                                                                                                                  \\\hline
       &  & \multicolumn{6}{c|}{OR, NIPW, AIPW}                                                  & \multicolumn{5}{c|}{OR, NIPW, AIPW}                                   & \multicolumn{5}{c|}{DML, TMLE}                                        \\
Method &  & \multicolumn{2}{l}{$\mu$} & \multicolumn{2}{l}{$e_1$} & \multicolumn{2}{l|}{$\mu_a$} & \multicolumn{2}{l}{$\mu$} & \multicolumn{2}{l}{$e_1$} & $\mu_a$      & \multicolumn{2}{l}{$\mu$} & \multicolumn{2}{l}{$e_1$} & $\mu_a$      \\\hline
L      &  & $\checkmark$      &       & $\checkmark$      &       & $\checkmark$       &        & $\checkmark$      &       & $\checkmark$      &       & $\checkmark$ & $\checkmark$      &       & $\checkmark$      &       & $\checkmark$ \\
M      &  & $\checkmark$      &       & $\checkmark$      &       & $\checkmark$       &        & $\checkmark$      &       & $\checkmark$      &       & $\checkmark$ & $\checkmark$      &       &                   &       &              \\
H2m   &  & $\checkmark$      &       & $\checkmark$      &       &                    &        & $\checkmark$      &       &                   &       &              & $\checkmark$      &       &                   &       &              \\

H1r  &  & $\checkmark$      &       & $\checkmark$      &       & $\checkmark$       &        & $\checkmark$      &       &                   &       &              & $\checkmark$      &       &                   &       &              \\
Hf2m   &  & $\checkmark$      &       &                   &       &                    &        & $\checkmark$      &       &                   &       &              & $\checkmark$      &       &                   &       &              \\

H2r  &  & $\checkmark$      &       & $\checkmark$      &       &                    &        & $\checkmark$      &       &                   &       &              & $\checkmark$      &       &                   &       &              \\
Hf1r     &  & $\checkmark$      &       &                   &       & $\checkmark$       &        & $\checkmark$      &       &                   &       &              & $\checkmark$      &       &                   &       &              \\
Hf2r  &  & $\checkmark$      &       &                   &       &                    &        & $\checkmark$      &       &                   &       &              & $\checkmark$      &       &                   &       &              \\
Mq     &  & $\checkmark$      &       & $\checkmark$      &       & $\checkmark$       &        & $\checkmark$      &       &                   &       &              & $\checkmark$      &       &                   &       &              \\
R      &  & $\checkmark$      &       & $\checkmark$      &       & $\checkmark$       &        & $\checkmark$      &       & $\checkmark$      &       & $\checkmark$ & $\checkmark$      &       & $\checkmark$      &       & $\checkmark$ \\
Rc     &  & $\checkmark$      &       & $\checkmark$      &       & $\checkmark$       &        & $\checkmark$      &       &                   &       &              & $\checkmark$      &       &                   &       &              \\
Rct    &  & $\checkmark$      &       & $\checkmark$      &       & $\checkmark$       &        & $\checkmark$      &       &                   &       &              & $\checkmark$      &       &                   &       &              \\
Rt     &  & $\checkmark$      &       & $\checkmark$      &       & $\checkmark$       &        & $\checkmark$      &       & $\checkmark$      &       & $\checkmark$ & $\checkmark$      &       &                   &       &              \\
S      &  & $\checkmark$      &       & $\checkmark$      &       & $\checkmark$       &        & $\checkmark$      &       & $\checkmark$      &       & $\checkmark$ & $\checkmark$      &       &                   &       &              \\
Gb      &  & $\checkmark$      &       & $\checkmark$      &       & $\checkmark$       &        & $\checkmark$      &       & $\checkmark$      &       & $\checkmark$ & $\checkmark$      &       & $\checkmark$      &       & $\checkmark$ \\
Gbt     &  & $\checkmark$      &       & $\checkmark$      &       & $\checkmark$       &        & $\checkmark$      &       & $\checkmark$      &       & $\checkmark$ & $\checkmark$      &       &                   &       &             \\\hline
\end{tabular}
\caption{Statitical and machine learning techniques used to estimate nuisance paramters $\mu(\cdot), e_a(\cdot), \mu_a(\cdot)$.}
\label{tab:checked_models}
\end{table}
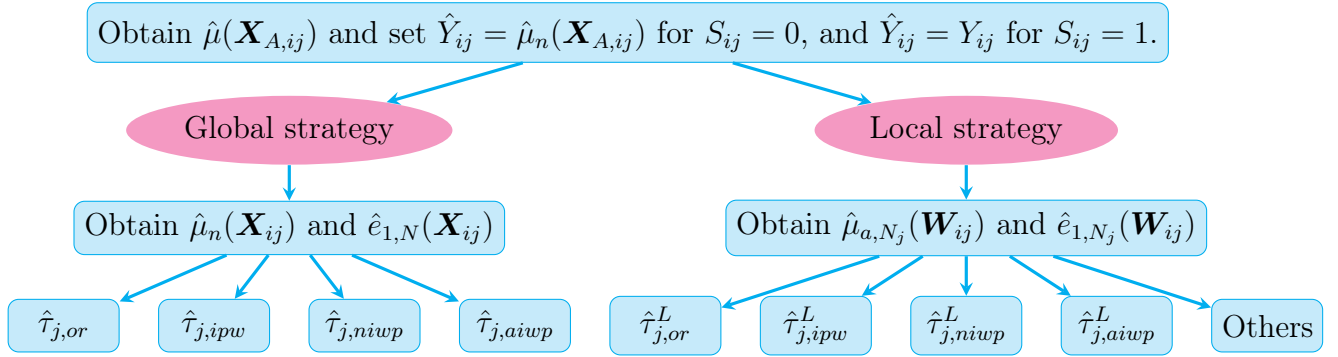
\begin{figure}[h] 
\centering 
\begin{tikzpicture}[node distance=2cm]
\node (title) [text centered, font=\large\bfseries] at (0,2) {Estimation strategies};
\node (process1) [process,  below of=title, yshift=1cm] {Obtain \( \hat{\mu}(\bm{X}_{A, ij}) \) and set \( \hat{Y}_{ij} = \hat{\mu}_n(\bm{X}_{A, ij}) \) for \( S_{ij} =0\), and \( \hat{Y}_{ij} = Y_{ij} \) for  \( S_{ij} =1\).};
\node (decision1) [decision, below of=process1, yshift=0.7cm,  xshift=-4.5cm] {Global strategy};
\node (decision2) [decision, below of=process1, yshift=0.7cm,  xshift=4.5cm] {Local strategy};

\node (process2) [process, below of=decision1, yshift=0.7cm] {Obtain $\hat{\mu}_{n}(\bm{X}_{ij})$ and $\hat{e}_{1,N}(\bm{X}_{ij})$};
\node (process3) [process, below of=decision2, yshift=0.7cm] {Obtain $\hat{\mu}_{a, N_j}(\bm{W}_{ij})$ and $\hat{e}_{1,N_j}(\bm{W}_{ij})$};

\node (process4) [stop, below of=process2, xshift=-3cm, yshift=0.7cm] {$\hat{\tau}_{j, or}$};
\node (process5) [stop, below of=process2, xshift=-1cm, yshift=0.7cm] {$\hat{\tau}_{j, ipw}$};
\node (process6) [stop, below of=process2, xshift=1cm, yshift=0.7cm] {$\hat{\tau}_{j, niwp}$};
\node (process7) [stop, below of=process2,  xshift=3cm, yshift=0.7cm] {$\hat{\tau}_{j, aiwp}$};

\node (process8) [stop, below of=process3, xshift=-4cm,yshift=0.7cm] {$\hat{\tau}^L_{j, or}$};
\node (process9) [stop, below of=process3, xshift=-2cm,yshift=0.7cm] {$\hat{\tau}^L_{j, ipw}$};
\node (process10) [stop, below of=process3, yshift=0.7cm] {$\hat{\tau}^L_{j, niwp}$};
\node (process11) [stop, below of=process3,  xshift=2cm, yshift=0.7cm]  {$\hat{\tau}^L_{j, aiwp}$};
\node (process12) [stop, below of=process3, xshift=4cm,yshift=0.7cm]  {Others};

\draw [arrow] (process1) -- (decision1);
\draw [arrow] (process1) -- (decision2);

\draw [arrow] (decision1) -- (process2);
\draw [arrow] (decision2) -- (process3);

\draw [arrow] (process2) -- (process4);
\draw [arrow] (process2) -- (process5);
\draw [arrow] (process2) -- (process6);
\draw [arrow] (process2) -- (process7);

\draw [arrow] (process3) -- (process8);
\draw [arrow] (process3) -- (process9);
\draw [arrow] (process3) -- (process10);
\draw [arrow] (process3) -- (process11);

\draw [arrow] (process3) -- (process12);

\end{tikzpicture}
\caption{Diagram of estimation strategies.}
	\label{fig:EstimationStrategies}
 \end{figure}

\begin{table}[]
\addtolength{\tabcolsep}{-0.28em}
\begin{tabular}{|llll|lllll|llllll|}\hline
\multicolumn{15}{|c|}{Global estimators}        \\\hline
\multicolumn{4}{|c|}{CSAE-OR}        & \multicolumn{5}{|c|}{CSAE-NIPW}            & \multicolumn{6}{c|}{CSAE -AIPW}                      \\
$\mu_a$ & MSE   & \% err. & Bias  & $\mu$ & $e_1$ & MSE   & \% err. & Bias  & $\mu$ & $e_1$ & $ \mu_a$ & MSE   & \% err. & Bias  \\\hline
H1r   & 0.120 & 19.122   & 0.100 & H2r & H1r & 0.115 & 13.561   & 0.101 & \textbf{H2r} & \textbf{X}     & \textbf{M}        & \textbf{0.107} & \textbf{5.659}    & 0.083 \\
X       & 0.166 & 64.788   & 0.053 & H2r & L     & 0.115 & 13.570   & 0.101 & H2r & S     & M        & 0.108 & 7.061    & 0.085 \\
Rt      & 0.173 & 71.513   & 0.067 & H2r & Xt    & 0.115 & 13.582   & 0.101 & H2r & X     & Rc       & 0.109 & 7.793    & 0.089 \\
R       & 0.173 & 71.521   & 0.067 & H2r & X     & 0.115 & 13.590   & 0.101 & H2r & X     & Rct      & 0.109 & 7.902    & 0.089 \\
Rct     & 0.174 & 72.261   & 0.063 & H2r & Rc    & 0.115 & 13.593   & 0.100 & H2r & X     & R        & 0.109 & 8.264    & 0.090\\\hline
\end{tabular}
\caption{5 best performing estimators under the global estimation strategies. MSE, mean squared error; \% err, increase of MSE in percentage with respect to the best performing method; Bias.}
\label{tab:best_global}
\end{table}

Tables~\ref{tab:best_global}-\ref{tab:best_local_2} present $5$-best performing strategies across all global and local estimation strategies. We can clearly see that among them, the best results were obtained by the global CSAE-AIPW (third column in Table~\ref{tab:best_global}) and local DML estimator (first columns in Table~\ref{tab:best_local_2}). These were further explored in Section 6 in the main document. 

\begin{table}[]
\addtolength{\tabcolsep}{-0.3em}
\begin{tabular}{|lllll|lllll|llllll|}\hline
\multicolumn{16}{|c|}{Local estimators}                                                                                                       \\\hline
\multicolumn{5}{|c}{CSAE-OR}                & \multicolumn{5}{|c}{CSAE-NIPW}            & \multicolumn{6}{|c|}{CSAE -AIPW}                      \\
$\mu$ & $\mu_a$ & MSE   & \% err. & Bias  & $\mu$ & $e_1$ & MSE   & \% err. & Bias  & $\mu$ & $e_1$ & $ \mu_a$ & MSE   & \% err. & Bias  \\\hline
H2r & M       & 0.120 & 19.122   & 0.100 & H2r & L     & 0.112 & 10.651   & 0.101 & H2r & Xt    & M        & 0.111 & 10.297   & 0.099 \\
H2r & L       & 0.166 & 64.788   & 0.053 & H2r & M     & 0.112 & 10.726   & 0.101 & H2r & X     & M        & 0.111 & 10.299   & 0.100 \\
H2r & Rt      & 0.173 & 71.513   & 0.067 & H2r & Rc    & 0.112 & 10.947   & 0.101 & H2r & Rc    & L        & 0.111 & 10.326   & 0.099 \\
H2r & R       & 0.173 & 71.521   & 0.067 & H2r & Rt    & 0.112 & 10.975   & 0.101 & H2r & Rt    & L        & 0.111 & 10.345   & 0.099 \\
H2r & X       & 0.174 & 72.261   & 0.063 & H2r & Xt    & 0.112 & 11.331   & 0.100 & H2r & M     & M        & 0.111 & 10.380   & 0.100\\\hline
\end{tabular}
\caption{5 best performing estimators under the local estimation strategies. MSE, mean squared error; \% err, increase of MSE in percentage with respect to the best performing method; Bias.}
\label{tab:best_local}
\end{table}

\begin{table}[]
\begin{tabular}{|llllll|llllll|}\hline
\multicolumn{12}{|c|}{Local estimators}                                                                    \\\hline
\multicolumn{6}{|c}{DML}                             & \multicolumn{6}{|c|}{TMLE}                           \\
$\mu$ & $e_1$ & $\mu_a$ & MSE   & \% err. & Bias   & $\mu$ & $e_1$ & $\mu_a$ & MSE   & \% err. & Bias  \\\hline
\textbf{H2r} & \textbf{X}     & \textbf{X}       & \textbf{0.101} & \textbf{0.000 }   & -0.025 & H2r & S*    & S*      & 0.114 & 13.025   & 0.099 \\
\textbf{H2m } & \textbf{X}     & \textbf{X}       & \textbf{0.105} & \textbf{4.265}    & -0.029 & H2m  & S*    & S*      & 0.118 & 16.377   & 0.089 \\
H2r & L     & R       & 0.108 & 6.496    & 0.061  & Mq    & S*    & S*      & 0.149 & 47.453   & 0.108 \\
H2r & L     & X       & 0.110 & 8.967    & 0.065  & Rt    & S*    & S*      & 0.166 & 64.156   & 0.016 \\
H2m  & L     & R       & 0.112 & 11.035   & 0.057  & R     & S*    & S*      & 0.166 & 64.215   & 0.030\\\hline
\end{tabular}
\caption{5 best performing DML estimators and TML estimators. MSE, mean squared error; \% err, increase of MSE in percentage with respect to the best performing method; Bias; S* refers to super-learner with a limited library of learners, see Section 6 in the main document.}
\label{tab:best_local_2}
\end{table}

\section{Additional results from the case study}\label{sec:SM_data_applicaiton}

Based on data availability and expert knowledge, the following features were selected to adjust for confounding: sex (two levels: female/male), nationality (two levels: Italian/not Italian), age (continuous), marital status, education level, and weekly working hours of the household head (HH); the household size; the type of contract (only included in the model for the outcome variable); the region indicator variable; and the average fiscal income by province. A full description of these variables, sourced from the EU-SILC 2012 survey and the Census 2011, is provided in Table~\ref{table:variables}. All features were normalized prior to effect estimation.

\begin{table}[h!]
\centering
\begin{tabular}{|l|l|l|}\hline
\textbf{Variable} & \textbf{HH} &\textbf{Description} \\\hline
Sex & $\checkmark$ &Two levels: Male/Female \\
Nationality & $\checkmark$&Two levels: Italian/Not Italian \\
Age & $\checkmark$&Continuous \\
Marital status  & $\checkmark$&Four levels: Never married/Married/ \\
                         & &Separated or Divorced/Widowed \\
Education level  &$\checkmark$ &Four levels: Pre-primary, Primary + Lower Secondary, \\
                          & &Upper-secondary + Post-secondary non-tertiary, Tertiary \\
Type of contract &$\checkmark$ &Two levels: Permanent/Short-term \\
Household size & &Continuous \\
Region & &Six levels: Lombardy, Tuscany, Umbria, Marche, \\
       & &Campania, Sicily \\
AFI/province & &Continuous (standardized) \\\hline
\end{tabular}
\caption{Description of variables available from EU-SILC 2012 and Population Census 2011; HH -- covariate at the level of the head of household; AFI -- average fiscal income. }
\label{table:variables}
\end{table}

\begin{figure}[ht]
    \centering
        \centering
 \includegraphics[width=\textwidth]{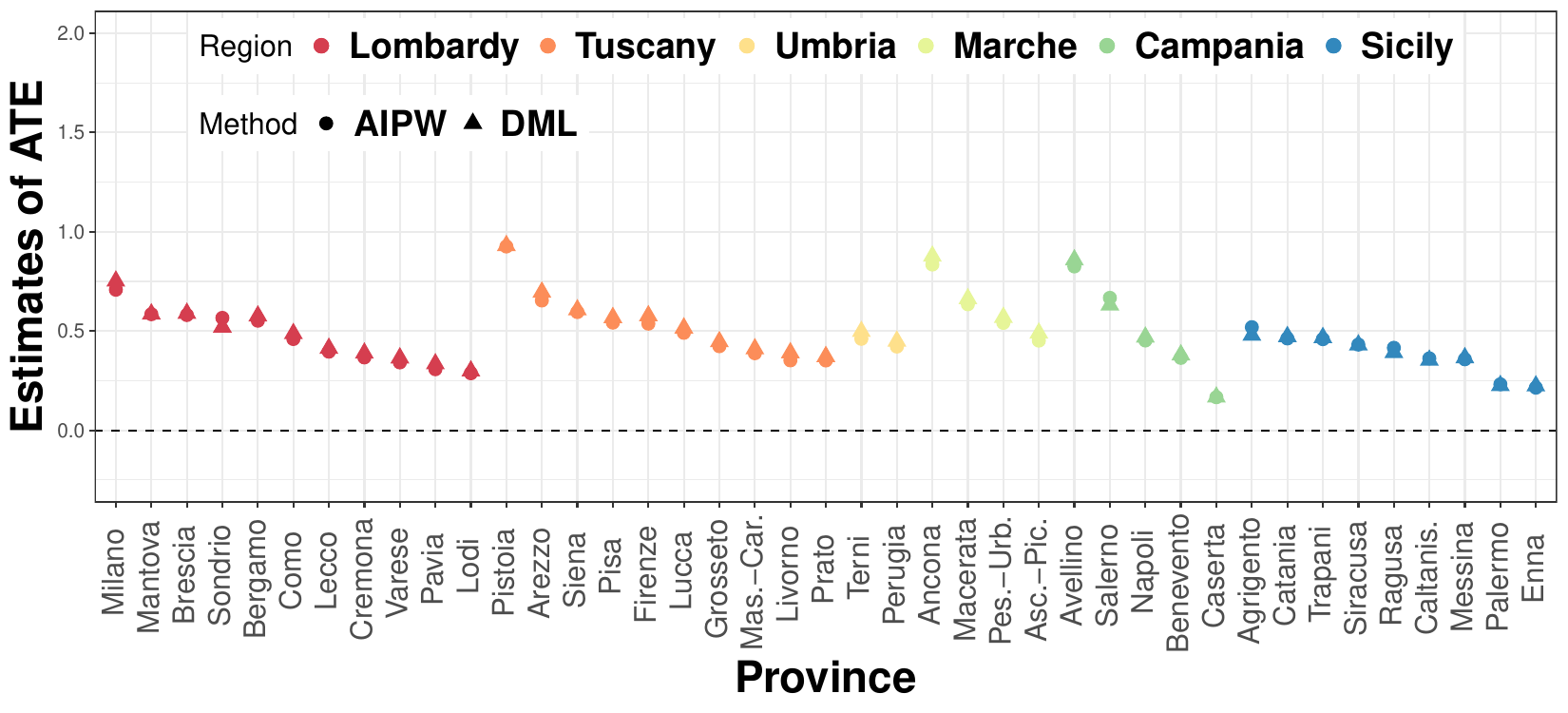}
        \label{fig:tau1}
     \caption{Point estimates of ATE obtained using global AIPW and local DML estmators.}
    \label{fig:estimates_DML_AIPW_map}
\end{figure}

Figure~\ref{fig:estimates_DML_AIPW_map} displays estimates of ATE on the original scale using  global CSAE-AIPW estimator and local DML estimator. As we can see, applying these two estimation strategies produced nearly identical estimates across provinces in Italy.

\bibliography{biblio}
\end{document}